\documentclass[conference]{IEEEtran}
\IEEEoverridecommandlockouts
\usepackage{cite}
\usepackage{amsmath,amssymb,amsfonts}
\usepackage{algorithmic}
\usepackage{graphicx}
\usepackage{textcomp}
\usepackage{tabularx} 
\usepackage{xcolor}
\usepackage{array}
\usepackage{subcaption}
\usepackage[table]{xcolor}
\usepackage{multirow}
\usepackage{graphicx}   
\usepackage[table]{xcolor}
\usepackage[hidelinks]{hyperref}
\def\BibTeX{{\rm B\kern-.05em{\sc i\kern-.025em b}\kern-.08em
    T\kern-.1667em\lower.7ex\hbox{E}\kern-.125emX}}

\usepackage{cite}

\usepackage{xcolor}

\begin{document}


\title{The xPU-athalon: Quantifying the Competition of \\AI Acceleration}

\author{\IEEEauthorblockN{Alicia Golden}
\IEEEauthorblockA{\textit{Harvard University} \\
Cambridge, MA}
\and
\IEEEauthorblockN{Carole-Jean Wu}
\IEEEauthorblockA{\textit{FAIR at Meta} \\
Cambridge, MA}
\and
\IEEEauthorblockN{Gu-Yeon Wei}
\IEEEauthorblockA{\textit{Harvard University} \\
Cambridge, MA}
\and
\IEEEauthorblockN{David Brooks}
\IEEEauthorblockA{\textit{Harvard University} \\
Cambridge, MA
}
} 

\maketitle

\begin{abstract}
The push for greater efficiency in AI computation has given rise to an array of accelerator architectures that increasingly challenge the GPU's long-standing dominance. In this work, we provide a quantitative view of this evolving landscape of AI accelerators, including the Cerebras CS-3, SambaNova SN-40, Groq, Gaudi, and TPUv5e platforms, and compare against both NVIDIA (A100, H100) and AMD (MI-300X) GPUs. We evaluate key trade-offs in latency, throughput, power consumption, and energy-efficiency across both (i) end-to-end workloads and (ii) benchmarks of individual computational primitives. Notably, we find the optimal hardware platform varies across batch size, sequence length, and model size, revealing a large underlying optimization space. Our analysis includes detailed power measurements across the prefill and decode phases of LLM inference, as well as quantification of the energy cost of communication. We additionally find that Cerebras, SambaNova, and Gaudi have 10-60\% higher idle power than NVIDIA and AMD GPUs, emphasizing the importance of high utilization in order to realize promised efficiency gains. Finally, we assess programmability across platforms based on our experiments with real profiled workloads, comparing the compilation times and software stack maturity required to achieve promised performance.

\end{abstract}

\IEEEpeerreviewmaketitle

\section{Introduction}

The recent explosion of large-scale AI has intensified the demand for efficient, scalable, and sustainable hardware systems. As AI workloads grow and efficiency pressures mount, hyperscalars are beginning to broaden their hardware portfolios in search of both capacity and architectural advantage. Recent announcements from major industry players point to a move towards multi-vendor deployments, including OpenAI’s plans for 6 GW of AMD GPU capacity~\cite{OpenAIAMDPartnership2025}. This shift has opened space to explore platforms that challenge the GPU's long-held dominance, reflecting a growing willingness to bet on alternative architectures that offer competitive performance or improved efficiency. 

In this context, a rapidly evolving ecosystem of AI accelerators has emerged, each built on a distinct architectural philosophy for delivering efficiency at scale. Novel accelerators aim to specialize computation, often trading general-purpose flexibility for higher performance~\cite{mlsysbook}. For example, wafer-scale architectures such as Cerebras exploit enormous amounts of on-chip SRAM to minimize data movement~\cite{cerebrasHotChips}. Deterministic hardware like Groq also rely on high-bandwidth SRAM, but pair it with a compiler-driven, instruction-stream architecture that delivers tightly scheduled execution~\cite{groqPaper, groqPaper2}. Meanwhile, other architectures such as SambaNova leverage spatially optimized data paths and kernel fusion to maximize performance~\cite{sambanovaPaper}.

\begin{figure}[t]
    \centering
    \includegraphics[width=0.65\linewidth]{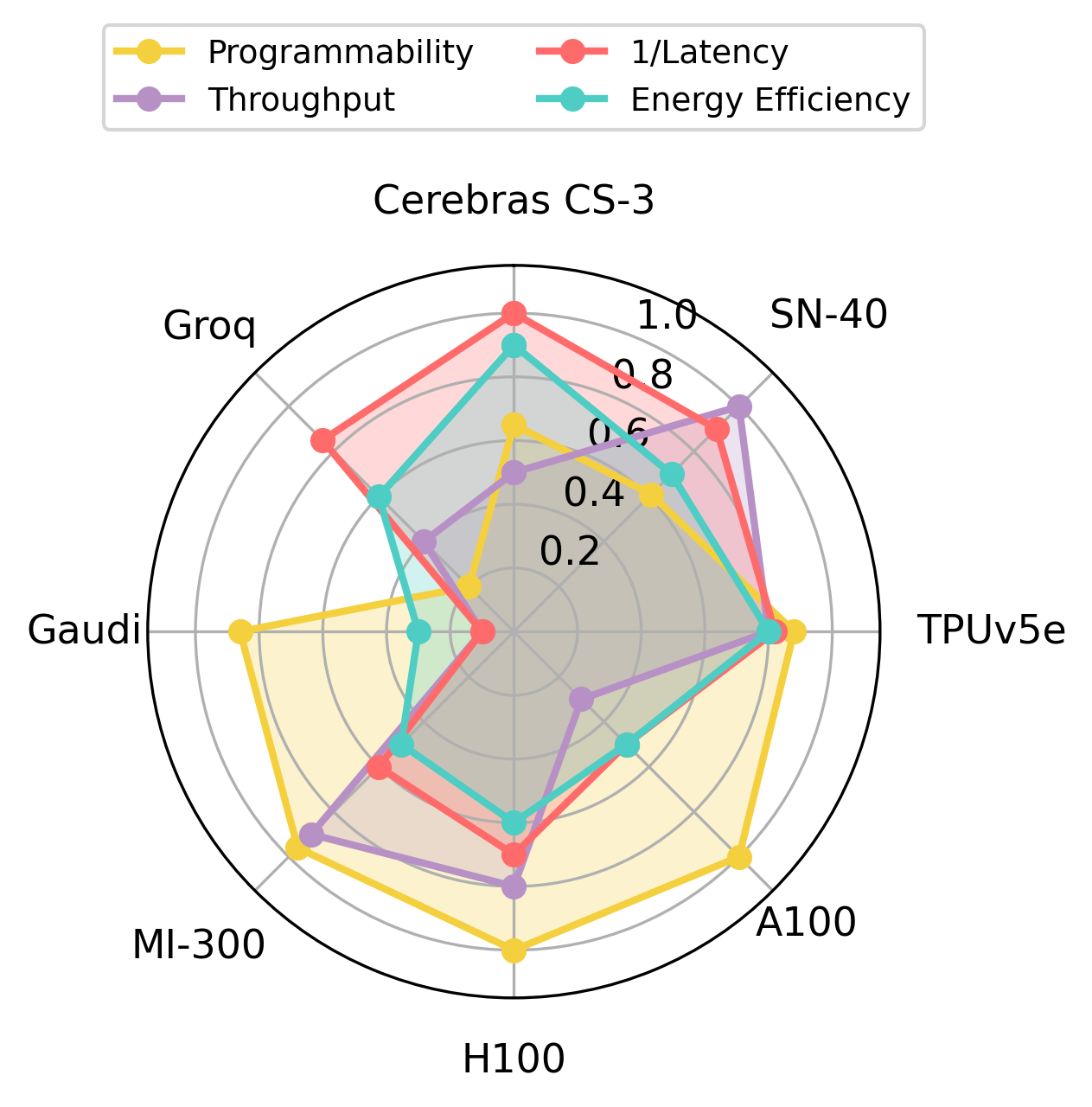}
    \caption{Overview of AI accelerator design space examined in this work. We characterize each platform along the axes of latency, energy-efficiency, and programmability. While Cerebras and Groq offer latency advantages, SambaNova sees benefits in high-throughput scenarios. We note that energy-efficiency gains are more subtle at scale, with Cerebras seeing benefits for small-scale out. Software-stack programmability also varies, with Gaudi and TPU leading, whereas SambaNova, Cerebras demonstrate stronger programmability than Groq. Software programmability is calculated via a combination of compilation time, tooling maturity, and kernel support.}
    \label{fig:spider_plot}
\end{figure}

\begin{figure*}[th]
    \centering

    \begin{subfigure}{0.40\linewidth}
        \centering
        \includegraphics[width=\linewidth]{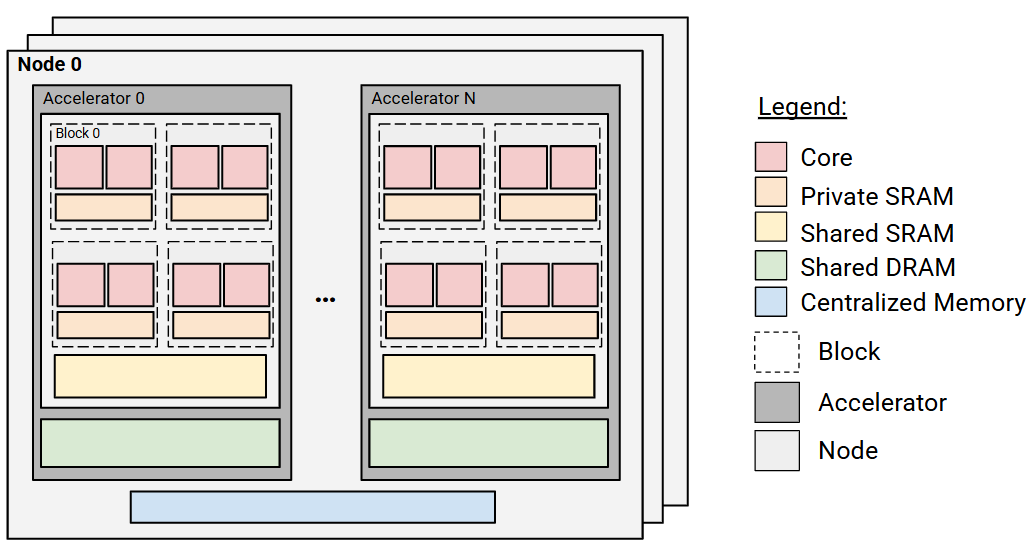}
        \caption{Generalized accelerator abstraction. Accelerators can be described at a primitive level as a collection of blocks, consisting of compute cores and private SRAM. These blocks are assembled with a shared on-chip SRAM and external DRAM to form an accelerator. Multiple accelerators form a node, which forms a system.}
        \label{fig:accel_architectures_a}
    \end{subfigure}
    \hfill
    \begin{subfigure}{0.58\linewidth}
        \centering
        \includegraphics[width=\linewidth]{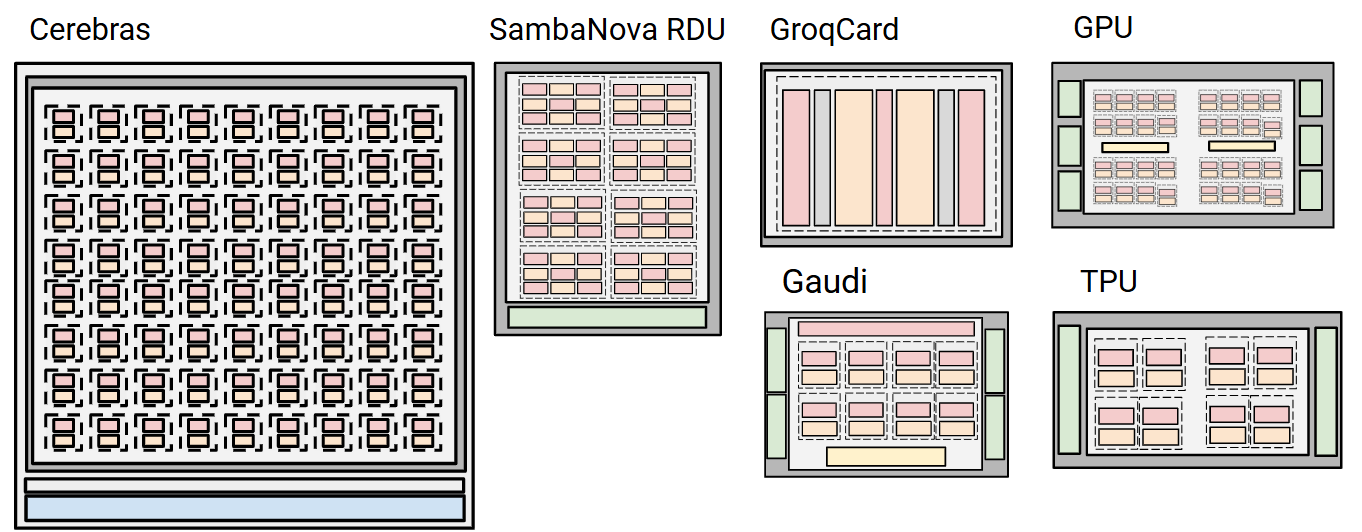}
        \caption{While all specific AI accelerators come with their own slate of terminology, each can be abstracted away into its primitive components. Here we show how each accelerator of interest maps to the generalized abstraction, including Cerebras (CS-3), SambaNova (SN-40), Groq, Gaudi, GPUs, and TPU, and compare the differing arrangements of memory and compute units. Note architectures not drawn to scale.}
        \label{fig:accel_architectures_b}
    \end{subfigure}

    \caption{Comparison of the suite of AI accelerator architectures examined in this work, grounded in a common abstraction.}
    \label{fig:accel_architectures}
\end{figure*}

However, determining the true advantages of these alternative hardware platforms is a challenging task. Figure \ref{fig:spider_plot} illustrates this open design space across eight AI hardware platforms, including Cerebras CS-3, SambaNova SN-40, Groq, Gaudi, TPUv5e, NVIDIA GPUs (A100, H100), and AMD MI-300. While Cerebras and Groq have latency advantages over more general-purpose GPUs at small-scale, SambaNova sees more optimal throughput for large batch-size scenarios. We note the energy efficiency benefits are nuanced, with Cerebras seeing reduced energy cost within a single wafer. The maturity of the software stack and programmability is also varied across platforms, with Gaudi and TPU having the most advanced stacks besides traditional GPUs, whereas SambaNova and Cerebras see clear advantages over the Groq stack. However, this only scratches the surface. To more fully understand the trade-offs that define this landscape, our work takes a disciplined look at the the architectural principles behind each accelerator and how they translate to performance.

In this paper, we present a quantitative analysis across hardware platforms, spanning both (i) end-to-end workload, and (ii) operator-level granularities. We begin by analyzing the upper bounds of each accelerator as determined by the hardware specifications. We then perform empirical evaluations directly on the target hardware platforms, measuring latency/token and probing primitive-level behavior through a series of microbenchmarks.

We dive deep into the architectural properties of each system, collecting measurements such as power traces across prefill and decode phases of LLM inference, and benchmarks to quantify the energy cost of communication. In doing so, we highlight key insights that contextualize AI performance across systems. Finally, we conclude with a discussion of challenges in accelerator adoption revealed by our experiments, including limitations in software-stack maturity.

The key contributions of this work are therefore as follows:

\begin{itemize}
    \item \textbf{We present a quantitative comparison across eight hardware platforms, including five novel AI accelerators and three GPU platforms}, and evaluate key trade-offs in latency, throughput, power consumption, and energy-efficiency across both (i) end-to-end workload, and (ii) operator-level granularities. Notably, we find the optimal hardware platform varies across batch size, sequence length, and model size, revealing a large underlying optimization space.
    
    \item \textbf{Power Efficiency.} We quantify power consumption across prefill and decode phases of LLM inference for each accelerator, and find that Cerebras, SambaNova, and AMD MI-300 utilize 75-100\% of TDP for memory-bound decode operations, while NVIDIA GPUs and Gaudi tend to use 45-60\% of TDP. We additionally observe 10-60\% higher idle power consumption of SambaNova, Cerebras and Gaudi relative to traditional GPUs, making high utilization critical for practical deployment.

    \item \textbf{Computational Primitives.} We evaluate a suite of computational primitives and find that Groq's latency advantage emerges primarily at small scales as opposed to scale-out scenarios, resulting in up to 300.16$\times$ speedup over H100 across primitive operations. 
    
    \item \textbf{Communication Energy.} We find that Cerebras significantly reduces the energy cost of communication as compared to H100 and Groq, since the wafer's large silicon area allows for more data to remain on-chip. 

    \item \textbf{Software Stack.} We identify programmability as a main bottleneck across accelerators: compilation times can reach 5000$\times$ higher for novel accelerators as compared to GPUs. The community should prioritize robust software development, especially compiler and kernel optimization, to unlock robust performance. 

\end{itemize}

\begin{figure}[t]
    \centering
        \centering
        \includegraphics[width=0.7\linewidth]{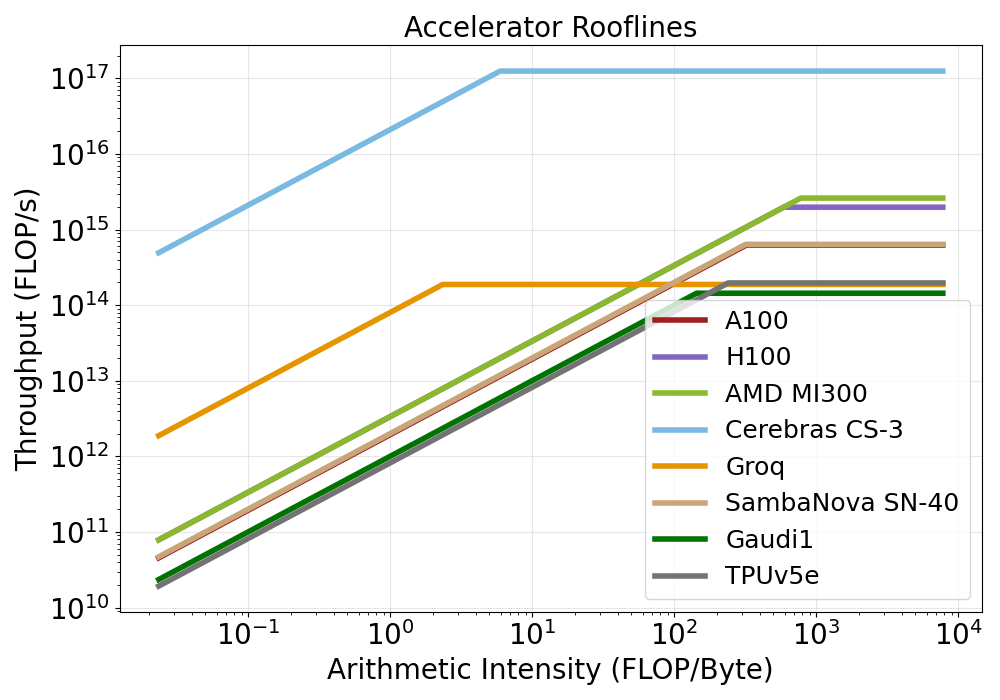}
        \caption{Roofline comparison across AI accelerator platforms. We observe that Cerebras CS-3 has roughly 2 orders of magnitude higher peak throughput than traditional GPU systems. Groq sees roughly 10$\times$ lower peak throughput than H100s, yet is less memory bandwidth bound due to its on-chip SRAM.}
        \label{fig:roofline}
\end{figure}

\section{Landscape of AI Accelerators} \label{ablationR}

\begin{figure*}[t]
    \centering
    \begin{subfigure}{0.32\linewidth}
        \centering
        \includegraphics[width=\linewidth]{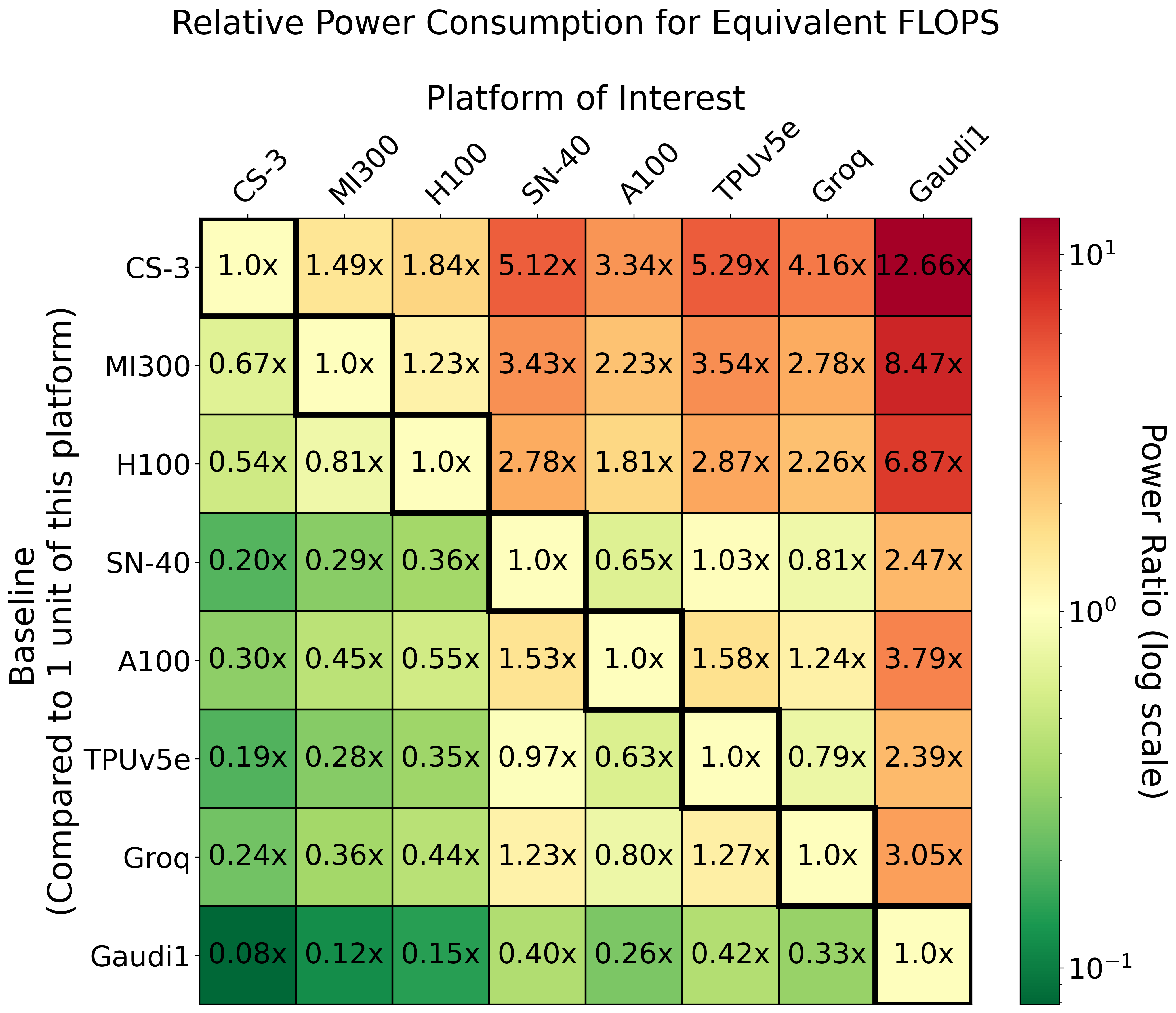}
        \caption{Relative power consumption for equivalent FLOPS. We plot the pairwise comparison of the power consumption for each combination of hardware platforms when scaled to equivalent number of FLOPS -- i.e., compared to 1 CS-3, MI300 scaled to the same number of FLOPS consumes 1.49$\times$ the power.}
        \label{fig:sub_power_equiv}
    \end{subfigure}
    \hfill
    \begin{subfigure}{0.32\linewidth}
        \centering
        \includegraphics[width=\linewidth]{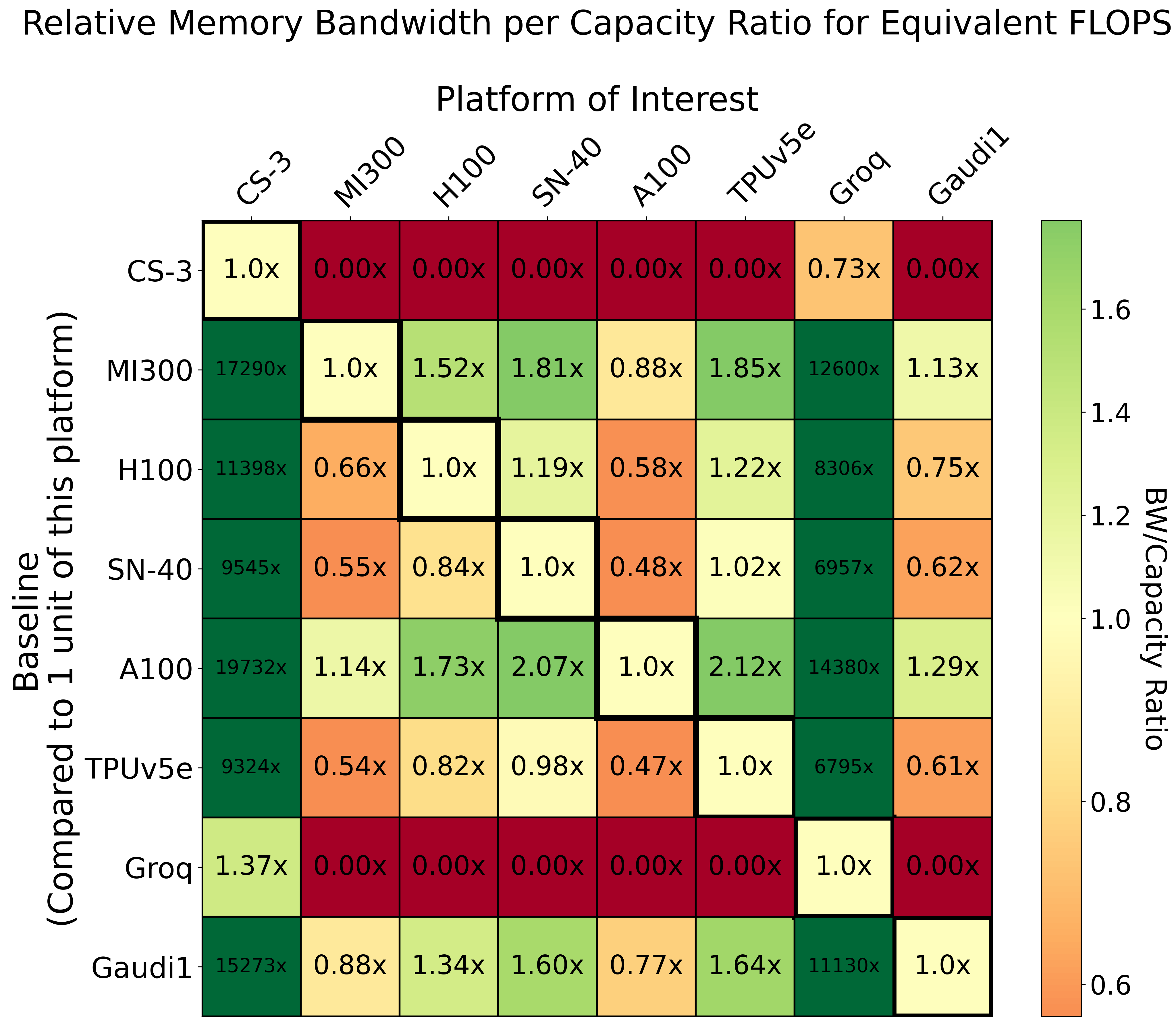}
        \caption{Relative memory bandwidth per capacity ratio for equivalent FLOPS highlights the trade-offs in memory systems, as introduced here ~\cite{rpu}. BW and capacity refer to accessible memory, i.e., working memory that holds model weights. CS-3 and Groq have more optimal BW/cap ratios due to use of SRAM.}
        \label{fig:sub_bandwidth_equiv}
    \end{subfigure}
    \hfill
    \begin{subfigure}{0.32\linewidth}
        \centering
        \includegraphics[width=\linewidth]{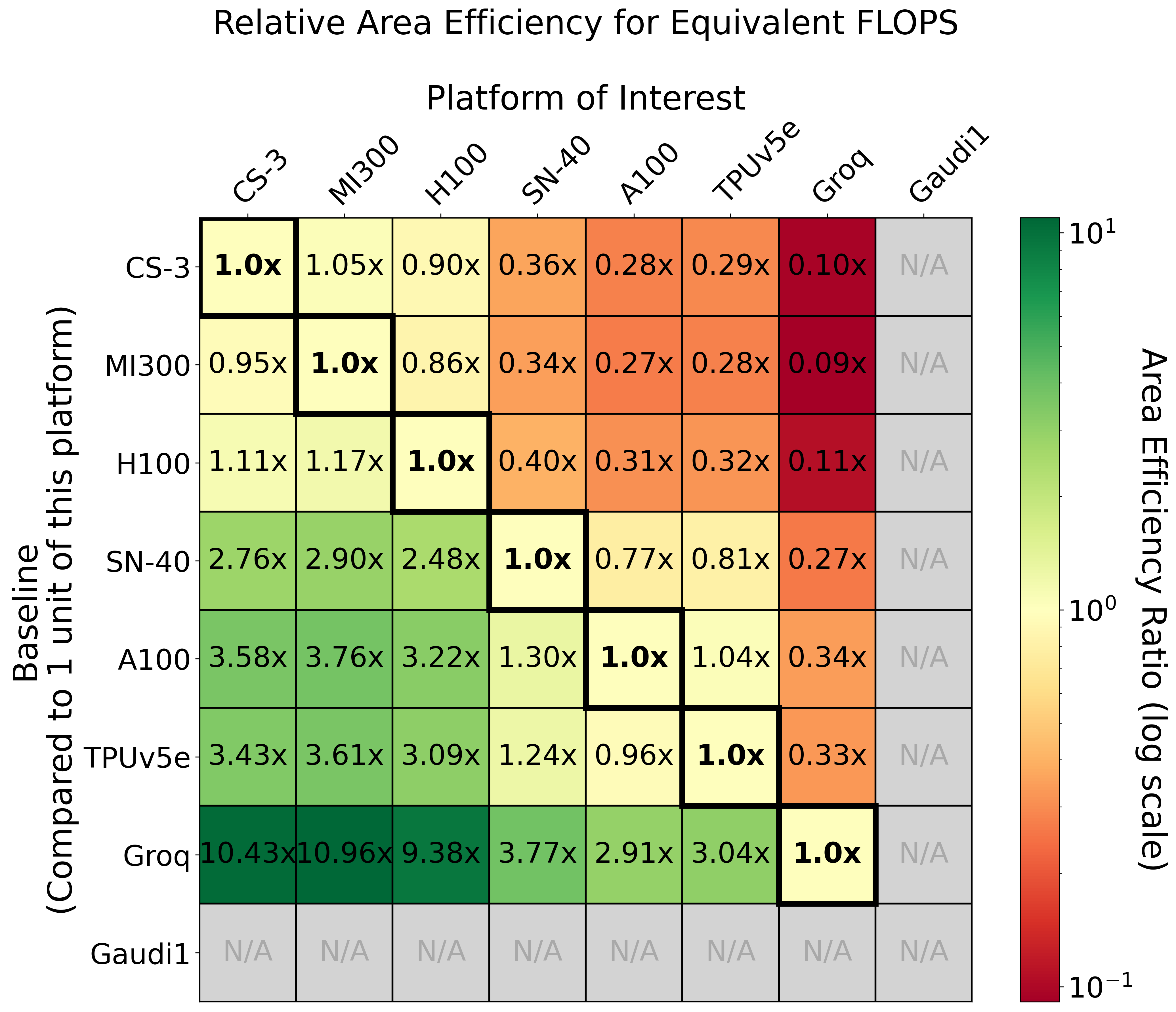}
        \caption{Relative area efficiency for equivalent compute ability. Higher values indicate more FLOPS per unit area ($mm^2$). We find that area efficiency of CS-3 is relatively comparable to H100 and MI300, while SN-40 and Groq have 0.40$\times$ and 0.11$\times$ the area efficiency of H100, respectively.}
        \label{fig:sub_bandwidth_equiv}
    \end{subfigure}
    \caption{Comparison across AI accelerator platforms for (a) power consumption and (b) memory bandwidth to capacity ratio, and (c) area efficiency each normalized to equivalent FLOPS.}
    \label{fig:platform_equivalency}
\end{figure*}

We first present an overview of the AI accelerator landscape, examining a range of hardware platforms including Cerebras, SambaNova, Groq, Gaudi, and TPU accelerators in comparison to NVIDIA and AMD GPUs. In this section, we compare key architectural characteristics of each platform, highlighting their distinctive design choices, capabilities, and trade-offs.

\subsection{Accelerator Architectures} Since each accelerator comes with its own slate of terminology, we begin by constructing a generalized abstraction from which we can examine each hardware platform. Figure \ref{fig:accel_architectures_a} shows our accelerator abstraction, with the smallest granularity being a \textit{block}, or combination of compute cores and private memory, specified by a core:memory ratio. These various blocks combine together into an \textit{accelerator}, where they share on-chip SRAM and off-chip DRAM capacity. Various numbers of accelerators come together to form a \textit{node} where they might have access to additional off-chip memory, and nodes are interconnected into a \textit{system}. Figure \ref{fig:accel_architectures_b} extends this abstraction to the many AI accelerators examined in this work, highlighting the corresponding compute and memory layouts for each architecture. We introduce each platform below:

\subsubsection{Cerebras}

Cerebras CS-3 is a wafer-scale processor -- a single, massive chip built from an entire silicon wafer~\cite{cerebrasHotChips}. Cerebras integrates a large amount of compute cores and memory directly on one piece of silicon, thereby eliminating many of the bottlenecks caused by chip-to-chip communication~\cite{cerebrasHotChips}.

\subsubsection{SambaNova}

SambaNova is a type of coarse-grained reconfigurable array, referred to as a Reconfigurable Data Unit (RDU)~\cite{sambanovaPaper}. Each RDU consists of a mesh of Pattern Memory Units (PMUs) and Pattern Compute Units (PCUs), which can be configured as either a systolic array or a SIMD core, depending on the workload~\cite{sambanovaPaper}. In this work, we evaluate the most recent SN-40 generation, although we revert to SN-30 for power measurements since it is the only platform we have physical access to. The generations differ mainly in their memory hierarchies but have similar TDP~\cite{sambanovaPaper, sambanovaServethehome}.

\subsubsection{Groq}

Often referred to as a Language Processing Unit (LPU), Groq is a specialized accelerator designed for LLM inference~\cite{groqPaper}. In contrast to traditional CPUs and GPUs, Groq has no memory hierarchy, only containing on-chip SRAM. The Groq chip is fully deterministic, reducing the need for control logic and congestion mitigation~\cite{groqPaper}.

\subsubsection{Gaudi}

The Gaudi v1 architecture integrates eight tensor processing cores (TPC), which serve as its primary compute units. Each TPC is a fully programmable VLIW (Very Long Instruction Word) processor optimized for matrix and tensor operations~\cite{gaudi}. While newer generations of Gaudi exist, only Gaudi1 is available on AWS public cloud~\cite{aws2024dl1}.

\subsubsection{TPUv5e}
The TPU is Google's custom-built processor design specifically to accelerate execution of large-scale AI models~\cite{tpuArch}. Built as a systolic array, TPU leverages tensor cores to achieve fast matrix multiplication, prioritizing HBM and scale-out in pod configurations. We evaluate TPUv5e in this context, since it is optimized for inference~\cite{googlecloud2025tpuv5e}.

\subsection{Comparing Architectural Implications}

We now provide an overview of how these architectural design choices translate to performance trade-offs.

Figure \ref{fig:roofline} presents a roofline model comparing the compute and memory capabilities of each platform, at the single-accelerator granularity of Figure \ref{fig:accel_architectures}. As shown, we observe that Cerebras CS-3 has roughly two orders of magnitude higher throughput (FLOP/s) than contemporary NVIDIA GPUs. We note both Groq and Cerebras substantially compress the memory-bound region of the roofline, a consequence of on-chip SRAM capacity and high effective memory bandwidth. 

However, it is important to note that this single accelerator roofline is not an apples-to-apples comparison across platforms. Rather, each accelerator operates at a fundamentally different order of magnitude, skewing direct comparisons. To address this and more fully understand the design space, Figure \ref{fig:platform_equivalency} highlights the comparison across each platform for an equivalent number of FLOPS. We plot the pairwise comparison of power consumption, memory bandwidth to capacity ratio, and silicon area for each combination of hardware platforms when scaled to the same number of FLOPS.

Figure \ref{fig:platform_equivalency}a highlights relative power consumption, assuming peak FLOPS and TDP. We find that CS-3 tends to be the most power efficient since compared to 1 H100, a CS-3 of equivalent FLOPS would require only 0.54$\times$ the power. MI300 is also emerges as power-efficient, consuming 0.81$\times$ the power of an H100 for an equivalent number of FLOPS. 

We repeat this analysis to compare the memory bandwidth/capacity ratio across platforms, a metric that gives insights into the overprovisoning of the memory system~\cite{rpu}. We analyze the bandwidth and capacity of the accessible memory, defined to be the working memory set that holds model weights (i.e., SRAM for CS-3 and Groq, DRAM for other platforms). We observe CS-3 and Groq have most optimal bandwidth to compute ratio due to their use of SRAM. We then examine area efficiency in Figure \ref{fig:platform_equivalency}c. We find that Groq has poorest area efficiency, with 10.43$\times$ the amount of area used to get equivalent FLOPS to CS-3. We now ground our analysis in a real-world use-cases, including LLM inference.

\section{Measurement Methodology}

\begin{table}[t]
\centering
\footnotesize
\renewcommand{\arraystretch}{0.95}
\setlength{\tabcolsep}{2pt}

\newcommand{\bigcheck}{{\scriptsize\color{green!60!black}\checkmark}}
\newcommand{\bigx}{{\scriptsize\color{red}$\times$}}

\begin{tabular}{|c|*{7}{c|}}
\hline
\textbf{Metric} &
\rotatebox{90}{A/H100} &
\rotatebox{90}{AMD MI300} &
\rotatebox{90}{Cerebras} &
\rotatebox{90}{SambaNova} &
\rotatebox{90}{Groq} &
\rotatebox{90}{Gaudi} &
\rotatebox{90}{TPUv5e} \\
\hline\hline

End-to-End Latency
  & \bigcheck & \bigcheck & \bigcheck & \bigcheck & \bigcheck & \bigcheck & \bigcheck \\ \hline

Per-Kernel Latency
  & \bigcheck & \bigcheck & \bigx & \bigx & \bigx & \bigx & \bigcheck \\ \hline\hline

\shortstack[c]{Power\\\footnotesize\textit{(*varying granularities)}}
  & \bigcheck & \bigcheck & \bigcheck & \bigcheck & \bigcheck & \bigcheck & \bigx \\ \hline

Temperature
  & \bigcheck & \bigcheck & \bigx & \bigx & \bigcheck & \bigcheck & \bigx \\ \hline\hline

\shortstack[c]{Compute Utilization\\\footnotesize\textit{(*differing definitions)}}
  & \bigcheck & \bigcheck & \bigcheck & \bigcheck & \bigx & \bigcheck & \bigcheck \\ \hline

\shortstack[c]{Memory Capacity \\Utilization}
  & \bigcheck & \bigcheck & \bigx & \bigcheck & \bigcheck & \bigcheck & \bigcheck \\ \hline

\shortstack[c]{Memory BW Utilization\\\footnotesize\textit{(*differing definitions)}}
  & \bigx & \bigx & \bigx & \bigcheck & \bigcheck & \bigx & \bigcheck \\ \hline

Number of Mem Copies
  & \bigx & \bigx & \bigx & \bigx & \bigcheck & \bigx & \bigx \\ \hline \hline

PCIe Stats
  & \bigcheck & \bigcheck & \bigx & \bigx & \bigx & \bigx & \bigx \\ \hline

IO Stats
  & \bigcheck & \bigcheck & \bigx & \bigx & \bigcheck & \bigx & \bigx \\ \hline

\end{tabular}

\caption{Available profiling metrics across hardware platforms. The varying profiling frameworks and uneven support of each metric make system-level comparisons a challenging task. We note metrics such as compute and memory bw utilization may have different definitions across platforms.}
\label{profiling_metrics}
\end{table}

\begin{table}[t]
  \centering
  \renewcommand{\arraystretch}{1.10}
  \setlength{\tabcolsep}{2pt}
  \scriptsize

  \newcommand{\hw}{{\color{green!60!black}\checkmark}}
  \newcommand{\api}{{\color{yellow!90!black}API}}
  \newcommand{\nope}{{\color{red}$\times$}}

  \resizebox{\columnwidth}{!}{
  \begin{tabular}{|c|c|c|c|c|c|c|c|}
    \hline
    \textbf{Model} & \textbf{A/H} & \textbf{MI300} & \textbf{CS-3} &
    \textbf{SN-40} & \textbf{Groq} & \textbf{Gaudi} & \textbf{TPUv5e} \\
    \hline \hline

    Llama-2-7B-hf & \hw & \hw & \nope & \nope & \hw & \hw & \hw \\ \hline
    Llama-3.1-8B  & \hw & \hw & \hw  & \api\textcolor{yellow!95!black}{**} & \api & \hw & \hw \\ \hline
    Llama-3.1-70B & \hw & \hw & \api & \api & \api & \hw & \hw \\ \hline

  \end{tabular}
  }

  \caption{Supported Models Across Hardware Platforms. Green check means supported on our physical hardware platforms, while API indicates supported via online interface only, thus limiting our profiling ability. Note we do have physical access to SN-30, which help supplement SN-40 measurements.}
  \label{workloads}
\end{table}

In this section, we detail our methodology for profiling workloads on each hardware platform. Note that simple API querying is \textit{not sufficient} for our system performance characterization, as we require measurements on the underlying architectural properties of each machine. We therefore obtain access to each accelerator and perform on-device profiling.

\subsection{Hardware Evaluation Environment.} We first describe our setup for each platform, including the configuration, software stack, and profiling framework used.

\textit{GPUs.} Our GPU experiments span both NVIDIA and AMD. For NVIDIA systems, we evaluate A100 and H100 GPUs deployed in multi-GPU server nodes (8 GPUs per DGX-Box) with NVLink interconnects, scaling out when appropriate. For AMD GPUs, we profile MI300 in a similar 8-GPU setup. We use nvidia-smi and pytorch profiler on NVIDIA GPUs, and leverage amd-smi and rocm-smi on AMD GPUs.

\textit{Cerebras.} We gained physical access to one CS-3 wafer, which we leverage to conduct experiments. We utilize cszoo and appliance version 2.5 to launch workloads, design custom benchmarks with cstorch language. We additionally control data movement through Cerebras Software Language (CSL) SDK ver 1.4. For power measurements, we record usage of the corresponding power distribution units (PDUs). Each CS-3 system contains 9 PDUs, which we sum to get total power.

\textit{SambaNova.} While we have API access to SN-40, collecting power measurements and running benchmarks requires access to physical hardware. For power and benchmark measurements, we thus revert to the SN-30 racks we have access to, and note the main difference between the two is in memory hierarchy. We utilize SambaFlow 1.24.1 to allow for workload execution on the SN-30 systems, and build upon the SambaNova Model Zoo to create custom scripts for benchmarking. We record power through the PDUs on the rack.

\textit{Groq.} We have access to 1 GroqRack (i.e., 72 GroqChips) for measurements, again running on the physical hardware platforms to collect system and compilation data. We then augment these measurements with reference API calls for more advanced workloads, due to limited model support on a single GroqRack. To design microbenchmarks we leverage tools such as Groqit and Groqflow 4.3.1 on the physical rack.

\textit{Gaudi.} We access Gaudi accelerators through AWS, which offers the first generation of the Gaudi machine. We utilize the Habana software stack and make use of CUDA's support for the HPU backend. We configure hl-smi tooling for profiling.

\textit{TPUv5e.} We access and profile the TPUv5e chip through Google Cloud Compute (GCP), leveraging the TPU VM instances. We measure reported latency, utilization, and other statistics through the tpu-info profiler.

\subsection{Profiling Tools.}

Since each accelerator platform comes with their own suite of tools, drawing comparisons across profiling metrics is difficult. Table \ref{profiling_metrics} summarizes the supported profiling information across each platform. All platforms provide end-to-end latency, but differ in kernel-level granularity: NVIDIA, AMD, Gaudi, and TPU expose per-kernel latencies, whereas Cerebras, SambaNova, and Groq abstract them away.

Across vendors, power and temperature measurement support varies. NVIDIA, AMD, and Gaudi expose built-in telemetry, while Cerebras and SambaNova require custom scripts and elevated privileges to query PDUs. Groq reports power via deterministic compiler output and a CLI. Measurement granularity differs widely (1 ms to 10 min), shaping our benchmark design. Utilization metrics also vary in availability and definition. Groq provides rich compilation statistics (e.g., per-hemisphere read/write activity), and some platforms expose PCIe and I/O metrics depending on system configuration.

\subsection{Workloads.}

Table \ref{workloads} summarizes a matrix of supported Llama models~\cite{llamaPaper} across all accelerators evaluated. We denote whether each workload can be run on our physical hardware system, via online API only, or not supported on either. We note that with API-only models, we cannot obtain low-level profiler statistics. We profile end-to-end inference workloads using standardized frameworks, such as vLLM~\cite{vllm} where applicable, to ensure consistency.

\subsection{Microbenchmark Suite} \label{microbenchmark_methodology}

We complement end-to-end profiling with a targeted microbenchmark suite that isolates accelerator limits. To construct this, we profile representative HuggingFace LLMs and gather a sweep of parameters such as sequence length, batch size, and hidden dimension. We extract the most frequent operators and their tensor shapes, yielding a focused set of benchmarks relevant to LLM inference.

\subsection{Measurement Collection} \label{power_methodology}

We profile power and utilization by instrumenting code with the relevant tools (e.g., nvidia-smi, amd-smi, tpu-info) and running background subprocesses during each workload or microbenchmark. We record idle power, peak power (typically by looping the benchmark to overcome coarse sampling), and transition periods. For end-to-end tasks like prefill/decode, we measure time to first token, pause, then measure per-token latency. To estimate energy, we use the observed peak power and assume it remains constant over the workload duration, compensating for profiler warmup and cooldown effects.

\section{Quantitative Comparison Across Hardware Platforms}

\begin{table*}[t]
  \centering
  \renewcommand{\arraystretch}{1.1}
  \setlength{\tabcolsep}{3pt}

  \resizebox{\textwidth}{!}{
  \begin{tabular}{|c|c|c|c|c|c|c|c|c|c|}
    \hline
    \cellcolor{gray!10}\textbf{Phase} &
    \cellcolor{gray!10}\textbf{Metric} &
    \cellcolor{gray!10}\textbf{A100}\cite{a100Datasheet} &
    \cellcolor{gray!10}\textbf{H100}\cite{h100Datasheet} &
    \cellcolor{gray!10}\textbf{AMD MI300}\cite{amdMI300Datasheet} &
    \cellcolor{gray!10}\textbf{Cerebras CS-3}\cite{cerebrasHotChips} &
    \cellcolor{gray!10}\textbf{SN-40}\cite{sambanovaDatasheet, sambanovaPaper} &
    \cellcolor{gray!10}\textbf{Groq}\cite{groqTokenomics} &
    \cellcolor{gray!10}\textbf{Gaudi1}\cite{gaudi} &
    \cellcolor{gray!10}\textbf{TPUv5e}\cite{googlecloud2025tpuv5e} \\
    \hline\hline

    Prefill & Compute (TFLOPS) & 624 & 1979 & 2614 & 125{,}000 & 638 & 188 & 144* & 197 \\
    \hline\hline

    \multirow{3}{*}{Decode}
     & Memory Type & DRAM & DRAM & DRAM & SRAM & DRAM & SRAM & DRAM & DRAM \\ \cline{2-10}
     & Capacity (GB) & 80 & 80 & 192 & 44 & 64 & 0.23 & 32 & 16 \\ \cline{2-10}
     & Bandwidth (B/s) & $1.935\times10^{12}$ & $3.35\times10^{12}$ & $5.3\times10^{12}$ & $2.1\times10^{16}$ & $2.0\times10^{12}$ & $8.0\times10^{13}$ & $1.0\times10^{12}$ & $819\times10^{9}$ \\ \cline{2-10}
    \hline
  \end{tabular}
  }

  \caption{Comparison of AI accelerator platforms in the context of LLM Inference. The prefill phase is compute-bound, and thus is heavily impacted by TFLOPS. In contrast, the decode phase is memory-bound, and we compare the accessible working memory of each accelerator here -- i.e., memory used to hold weights and KV-Cache in LLM inference.}
  \label{tab:hw_table}
\end{table*}

\subsection{Understanding the Optimization Space for LLM Inference} \label{theoretical}

We begin our quantitative comparison of AI accelerators with a characterization of end-to-end LLM inference. To do so, we first leverage established models to understand the optimization trade-off space, since our physical measurement set-ups are limited to particular workload configurations and do not uniformly allow for batch-size scaling.

Table \ref{tab:hw_table} situates our accelerator comparison in the context of the two key LLM inference phases: prefill, a compute-bound stage dominated by peak throughput, and decode, a memory bandwidth-bound phase. While performance can be estimated by assuming prefill is compute bound and decode is memory-bandwidth bound, this fails to account for the scale-out required based on memory capacity of the platform. In fact, to host a typical LLM workload such as Llama-3.1-70B~\cite{llamaPaper}, the amount of scale-out necessary varies from 2 -- 576 accelerators. This manifests in exposed communication, which can be significant depending on parallelization strategy. 

Therefore, we find that in order to fairly compare across systems, we must consider the \textit{distributed inference setting}. We follow established modeling procedures, using~\cite{flops2} to calculate the flop count for each pass through the transformer, and~\cite{kvCache} to calculate the memory capacity needed to hold the model weights and KV cache. Based on~\cite{cerebrasHotChips, groqParallelisms, sambanovaParallelisms}, we find that Cerebras uses pipeline parallelism, while Groq and SambaNova each use a combination of tensor parallelism and pipeline parallelism to execute Llama models. To calculate data transfer during pipeline parallelism, we use~\cite{narayanan2021efficient, deepspeedPPFormula}, and for tensor parallelism we follow~\cite{narayanan2021efficient, allReduceCommunicationFormula, tp2}. 

\begin{figure}[t]
    \centering
        \includegraphics[width=0.9\linewidth]{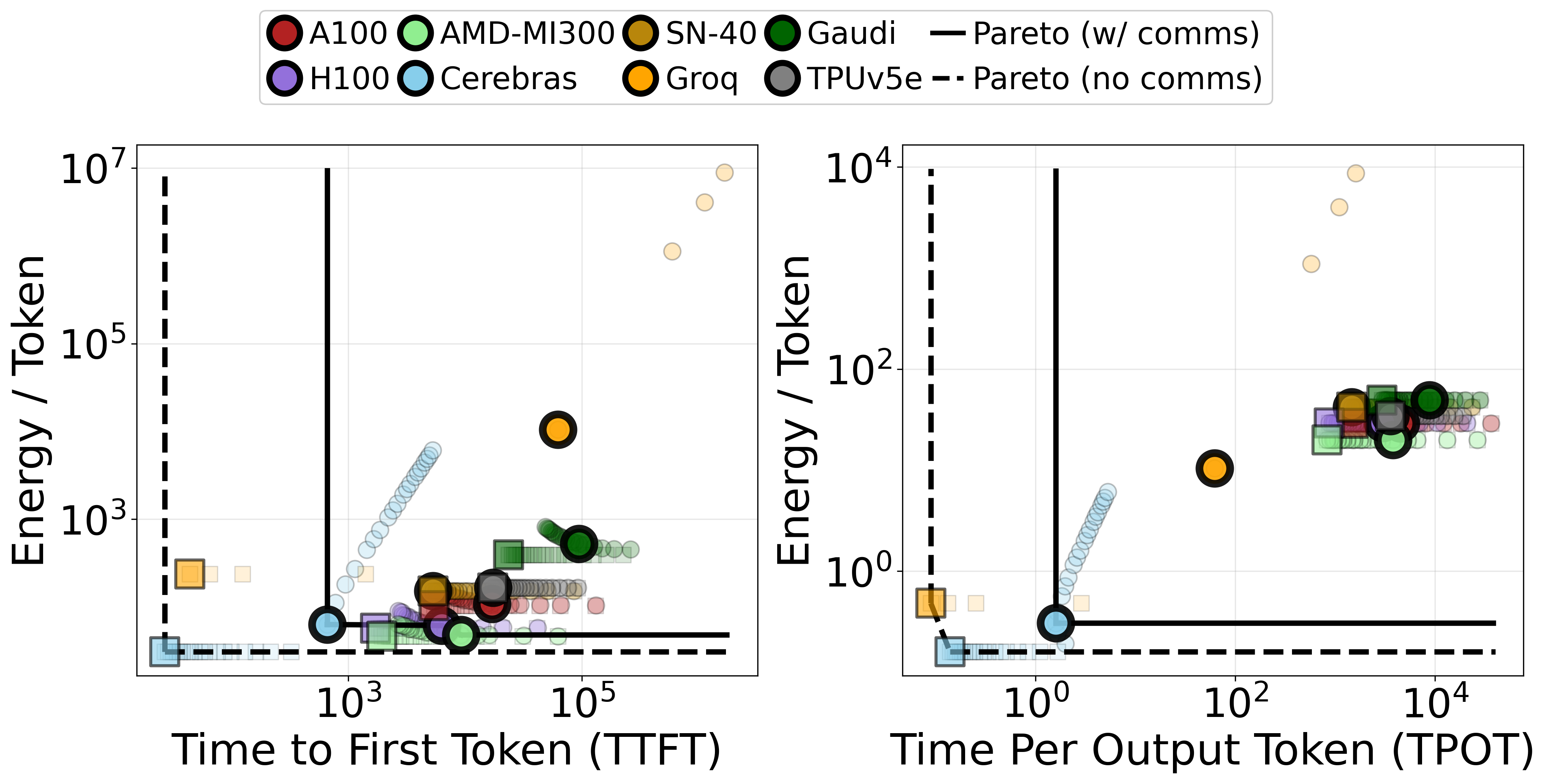}
    \caption{Analyzing theoretical bounds for distributed LLM inference. We show optimistic and realistic pareto-optimal curves for prefill (left) and decode (right), where squares show the case with zero exposed communication latency (optimal) and circles highlight realistic case with communication exposed for Llama-3.1-70B (bs 1, seqlen 128k). We find scale-out plays an important role in evaluating accelerators for a given workload --- e.g., Groq is on the decode pareto frontier when zero communication is exposed, but is no longer optimal when factoring in communication necessary for scale-out.}  \label{fig:prefill_decode}
\end{figure}

\begin{figure}[t]
    \centering
        \includegraphics[width=\linewidth]{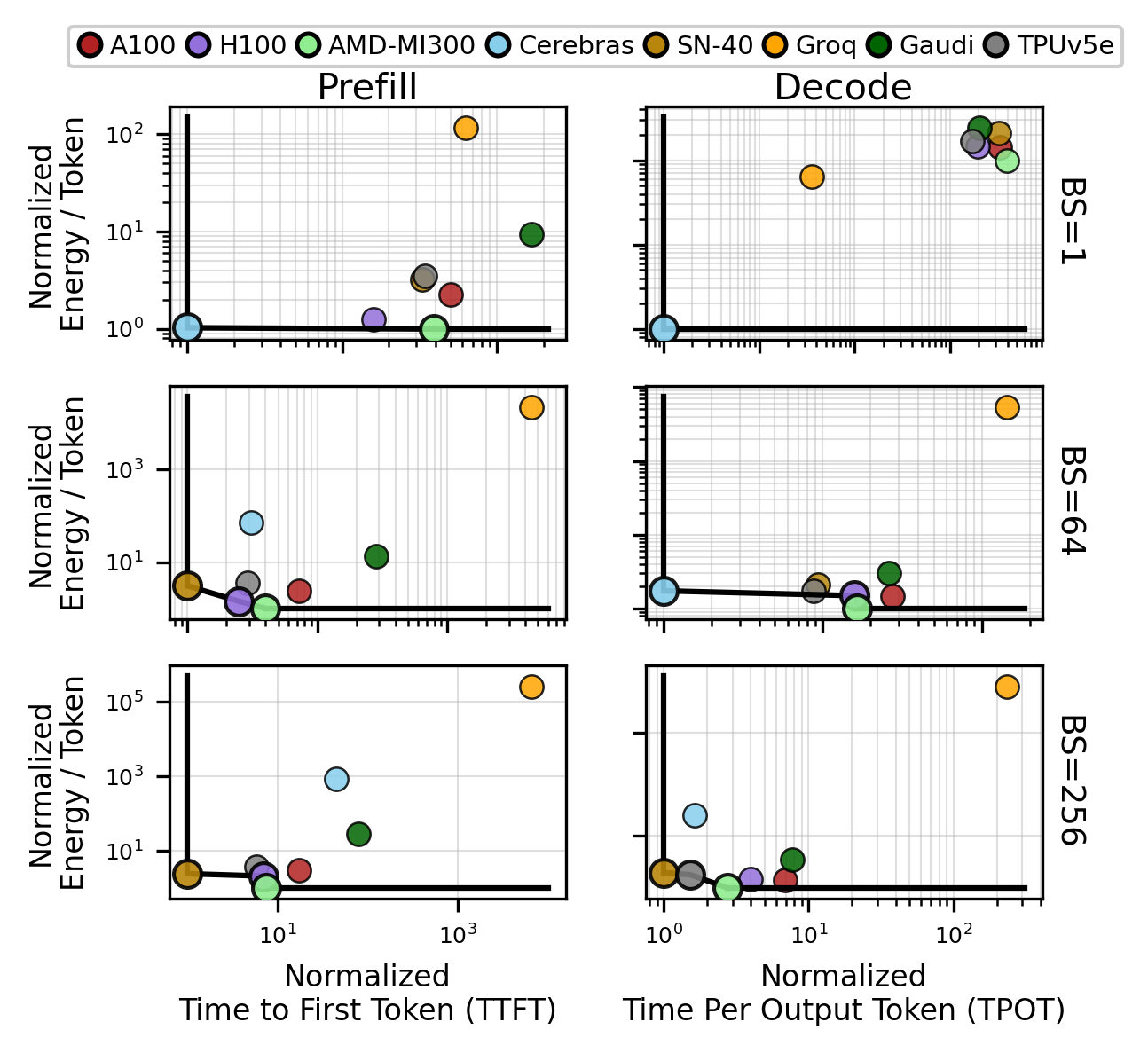}
    \caption{Optimal latency and energy trade-off depends heavily on batch size, sequence length, and number of parameters. Here we highlight a batch size sweep for Llama-3.1-70B, seqlen 128k. We find while Cerebras is optimal at low batch sizes, SambaNova, AMD-MI300, TPU, and H100 emerge on the pareto-frontier for more throughput-oriented scenarios.}  \label{fig:prefill_decode_2}
\end{figure}

We first examine how scale-out impacts performance trade-offs regarding latency and energy/token for the Llama-3.1-70B model. As shown in Figure \ref{fig:prefill_decode}, we examine the pareto-optimal curve for prefill (Time to First Token) and decode (Time Per Output Token) and compare to energy per token. We compute both optimistic and realistic scenarios, where square points show the best-case outcome of zero exposed communication latency, and circle points highlight a more realistic view of the exposed communication required for distributed inference. We sweep number of accelerators that satisfy the memory capacity threshold as well as the corresponding parallelism configurations. We find accounting for scale-out plays an important role in evaluating accelerators --- assuming zero exposed communication latency leads to different accelerators on the pareto-optimal curve. For example, Groq achieves pareto optimality on decode under zero-communication assumptions, yet loses that position when scale-out communication costs are incorporated. This highlights the drawback of Groq's highly disaggregated architecture when serving large models.

\vspace{0.5em}
\noindent
\fcolorbox{black!100}{gray!10}{
  \parbox{0.95\linewidth}{
    \vspace{0.2em}
    \underline{Key Takeaway 1:} When evaluating accelerators of different scales, accounting for distributed scale-out effects is essential: platforms that appear optimal under isolated assumptions (e.g., Groq) may lose their advantage once real communication overheads are considered.
    \vspace{0.2em}
  }
}
\vspace{0.5em}

However, Figure \ref{fig:prefill_decode} represents only a single view of a much larger optimization space --- and in practice, we find the optimal latency–energy trade-off depends heavily on batch size, sequence length, and model size. While we perform a comprehensive optimization space exploration, we show a representative example in Figure \ref{fig:prefill_decode_2} due to space constraints. We find that Cerebras is optimal in terms of energy latency trade-off at low batch sizes for both prefill and decode, but quickly drops off the Pareto curve as batch size grows. We note that Cerebras remains optimal for higher batch sizes in the decode phase as compared to prefill phase. We additionally find that at higher, throughput-oriented batch sizes, platforms such as SambaNova, MI300, H100, and TPU become Pareto-optimal. We note that the batch-size threshold at which each platform emerges on the curve also varies with sequence length and model size: for example, our exploration shows that CS-3, H100, and MI300 all appear on the pareto frontier for Llama-3.1-405B, even at batch size 1.

\vspace{0.5em}
\noindent
\fcolorbox{black!100}{gray!10}{
  \parbox{0.95\linewidth}{
    \vspace{0.2em}
    \underline{Key Takeaway 2:} Optimal LLM inference accelerators vary with batch size, sequence length, and model scale. For example, we find that while Cerebras performs best at small batch sizes for Llama-3.1-8B models at 128K sequence length, other platforms such as SambaNova, AMD MI300, TPU, and H100 move onto the Pareto frontier as throughput demands increase.
    \vspace{0.2em}
  }
}
\vspace{0.5em}

\begin{figure}[b]
    \centering
    \includegraphics[width=0.8\linewidth]{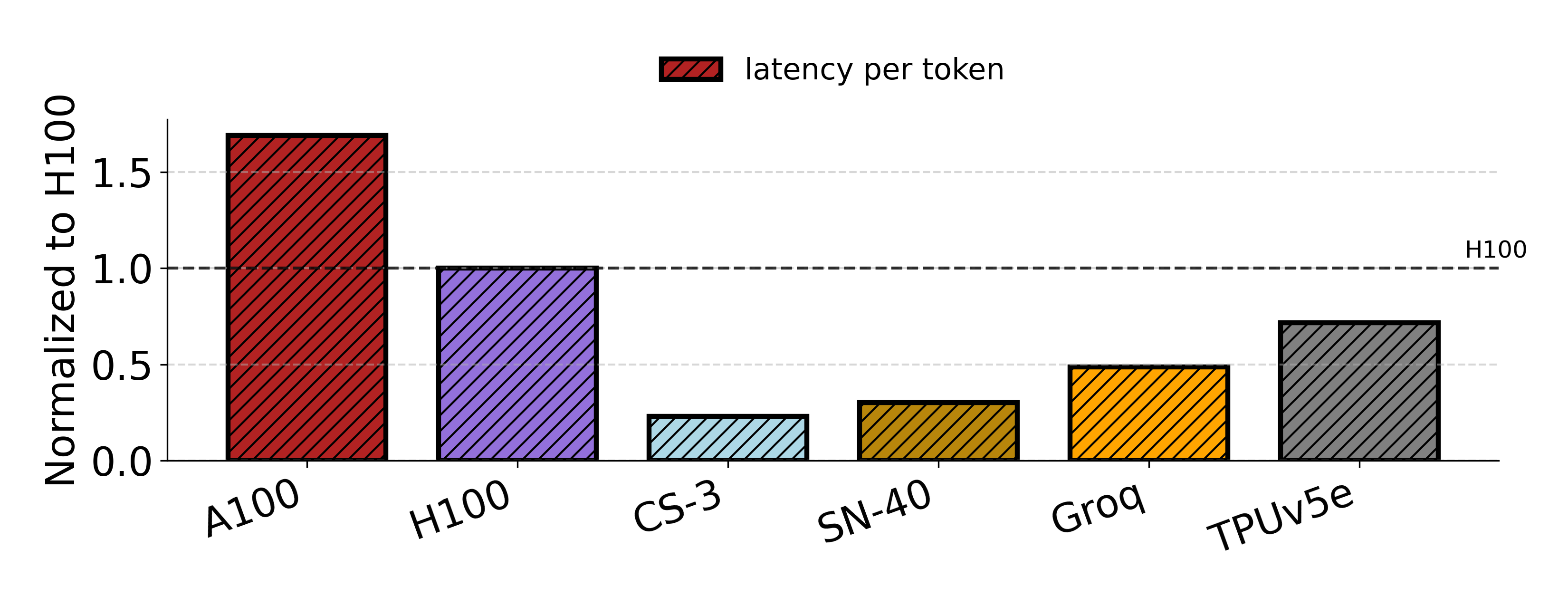}
    \caption{Overall latency comparison for low-batch inference across accelerators for Llama-3.1-8B. We find Cerebras has most optimal latency/token, at 22.89\% of H100 values.}
    \label{fig:accel_arena}
\end{figure}

We now highlight an example of low-batch inference with a small model available with our hardware constraints. Figure \ref{fig:accel_arena} shows overall latency per token comparison for Llama-3.1-8B, which we measure across physical platforms except for SN-40 and Groq which we leverage API (due to limited support, as shown in Table \ref{workloads}). We find Cerebras achieves lowest latency per token, at 22.89\% of baseline H100 values. For this small model and batch size, Groq and SN-40 see more optimal latency per token than H100s as well, at 30.03\% and 48.61\%.

While the observed performance gap across platforms is a result of both hardware factors and software maturity, we note hardware factors dominate. In particular, the hardware architectural properties of each platform (i.e., FLOPS, memory bandwidth) are the primary drivers. Take, for example, the gap between Groq and H100. Groq, which is a completely SRAM-based architecture, with a memory-bandwidth-per-capacity ratio that is orders of magnitude higher than H100’s DRAM architecture (see Figure 2). This alleviates the memory-bandwidth bottleneck found on traditional GPUs and enables the acceleration of LLM decode, which is a bandwidth bound workload.

In contrast and for context, we conduct a separate set of measurements to evaluate the impact of software changes on performance. For example, we collect Groq performance results over a six-month period of software updates that span 13 different package versions. We observe no statistically significant change in performance for the model configuration evaluated, indicating that software revisions over this interval do not meaningfully affect the measured results. Similarly, we conduct a set of ablation experiments on H100 in which we disable current state-of-the-art software optimization flags (such as Flash Attention3, vLLM optimization levels, etc.). We observe that removing these optimizations can increase latency per token by up to 2.14x, with the impact varying across model configurations and optimization types. We note that as software continues to evolve, additional incremental improvements may occur across platforms. We expect these gains to accrue more to platforms such as Groq that have comparatively less-mature stacks, which would modestly widen their advantage over H100, though hardware remains the dominant driver of the gap.

\subsection{Power Efficiency Across Platforms.}

To gain a deeper understanding of energy and power efficiency beyond the TDP proxy, we probe the power efficiency of each hardware platform. To do so, we capture power traces for full end-to-end model execution. We follow the power methodology outlined in Section \ref{power_methodology}, separately quantifying power for prefill and decode phases of LLM execution, allowing the system to return back to idle power in-between. 

\begin{figure}[t]
    \centering
    \includegraphics[width=\linewidth]{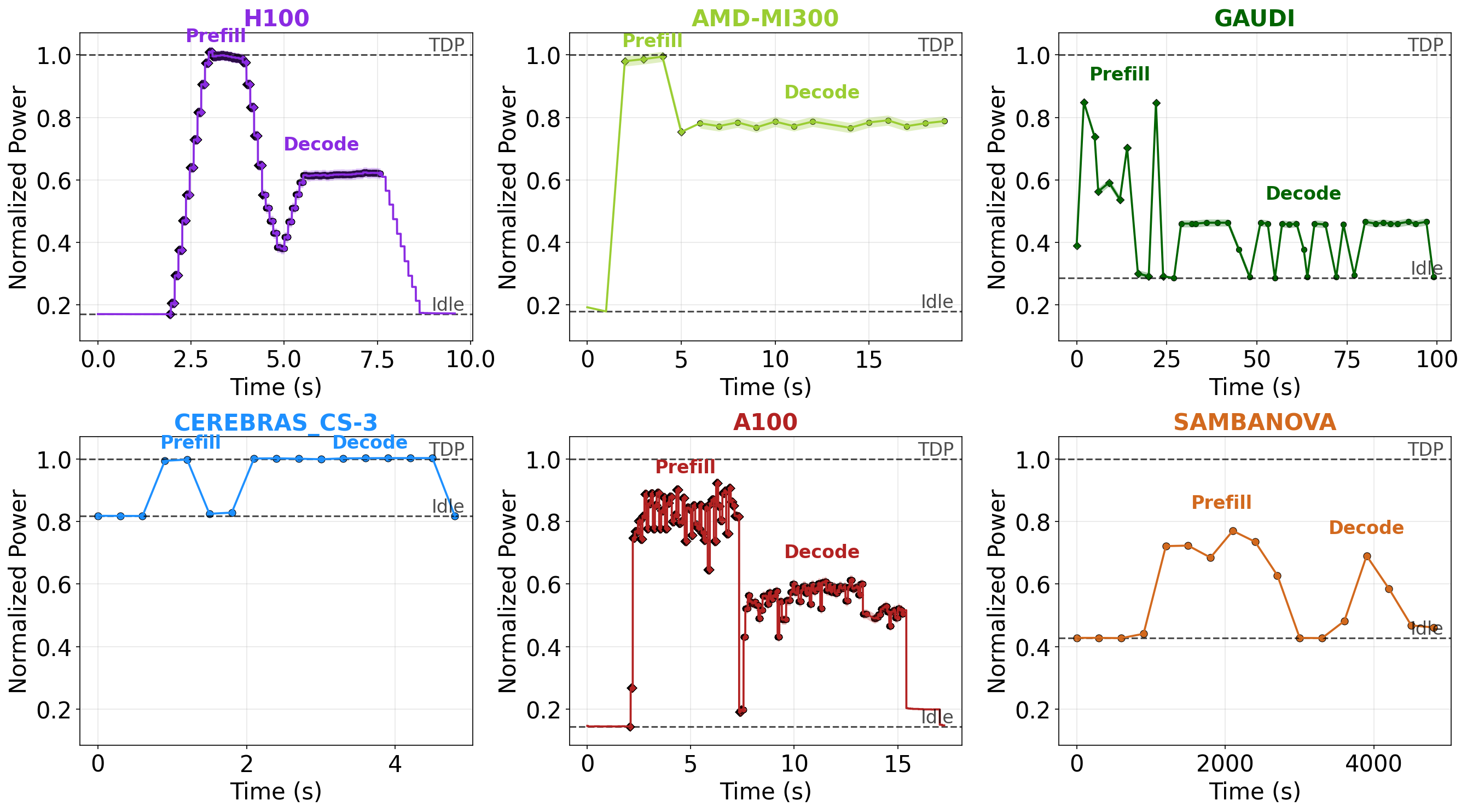}
    \caption{Power traces for Llama-3.1-8B during prefill and decode phases of execution. While Gaudi and NVIDIA GPUs utilize 45-60\% of TDP for decode, SambaNova and AMD MI300 utilize closer to 75-80\%. In contrast, Cerebras utilizes 100\% of TDP for both prefill and decode.}
    \label{fig:power_traces}
\end{figure}

Figure \ref{fig:power_traces} showcases power traces for Llama 3.1-8B inference across A100, H100, AMD-MI300, Cerebras CS-3, SambaNova, and Gaudi platforms. We note that while prefill consumes higher power consumption than decode stages, the \textit{relative difference} in power consumption between prefill and decode varies depending on the accelerator. In particular, prefill consumes 75-100\% of TDP consistently across platforms. For decode, NVIDIA GPUs such as A100 and H100 see decode power consumption at roughly 50-60\% of TDP. In contrast, SambaNova and AMD-MI300 see decode power consumption at 75\% and 80\% of TDP, respectively. This indicates these platforms have higher power consumption for a memory-bound decode stage, and may consume more power moving data in and out of memory. Cerebras decode power consumption is equal to prefill, suggesting memory-bound operations take as much power as compute-bound ones. 

\vspace{0.5em}
\noindent
\fcolorbox{black!100}{gray!10}{
  \parbox{0.95\linewidth}{
    \vspace{0.2em}
    \underline{Key Takeaway 3:} The relative difference in power consumption between prefill and decode varies across accelerators. While NVIDIA GPUs typically utilize 50-60\% of prefill power consumption for the decode phase, SambaNova and Cerebras utilize 75\% and 100\%, respectively. 
    \vspace{0.2em}
  }
}
\vspace{0.5em}

We additionally examine the relative idle power states of each platform as compared to power under high compute-bound workloads. We find idle power is 20\% of TDP across NVIDIA GPUs (A100 and H100) and AMD MI300. Gaudi has slightly higher idle power, at 30\% of TDP, meanwhile SambaNova idle power sits at 40\% of TDP. Cerebras has the highest idle power, at 80\% of TDP and note it reaches energy-per-token parity with a 32-GPU H100 cluster at 34\% duty cycle. We find that specialized hardware such as Cerebras, Gaudi, and SambaNova have higher idle power than more general-purpose platforms. This has important implications for utilization --- as these systems must be highly utilized in order to be efficient.

\vspace{0.5em}
\noindent
\fcolorbox{black!100}{gray!10}{
  \parbox{0.95\linewidth}{
    \vspace{0.2em}
    \underline{Key Takeaway 4:} The high idle power consumption of Cerebras, SambaNova, and Gaudi relative to traditional GPUs makes achieving high utilization critical for practical deployment. 
    \vspace{0.2em}
  }
}
\vspace{0.5em}

\begin{figure*}[hbt!]
    \centering
    \begin{subfigure}[b]{0.49\linewidth}
        \centering
        \includegraphics[width=\linewidth]{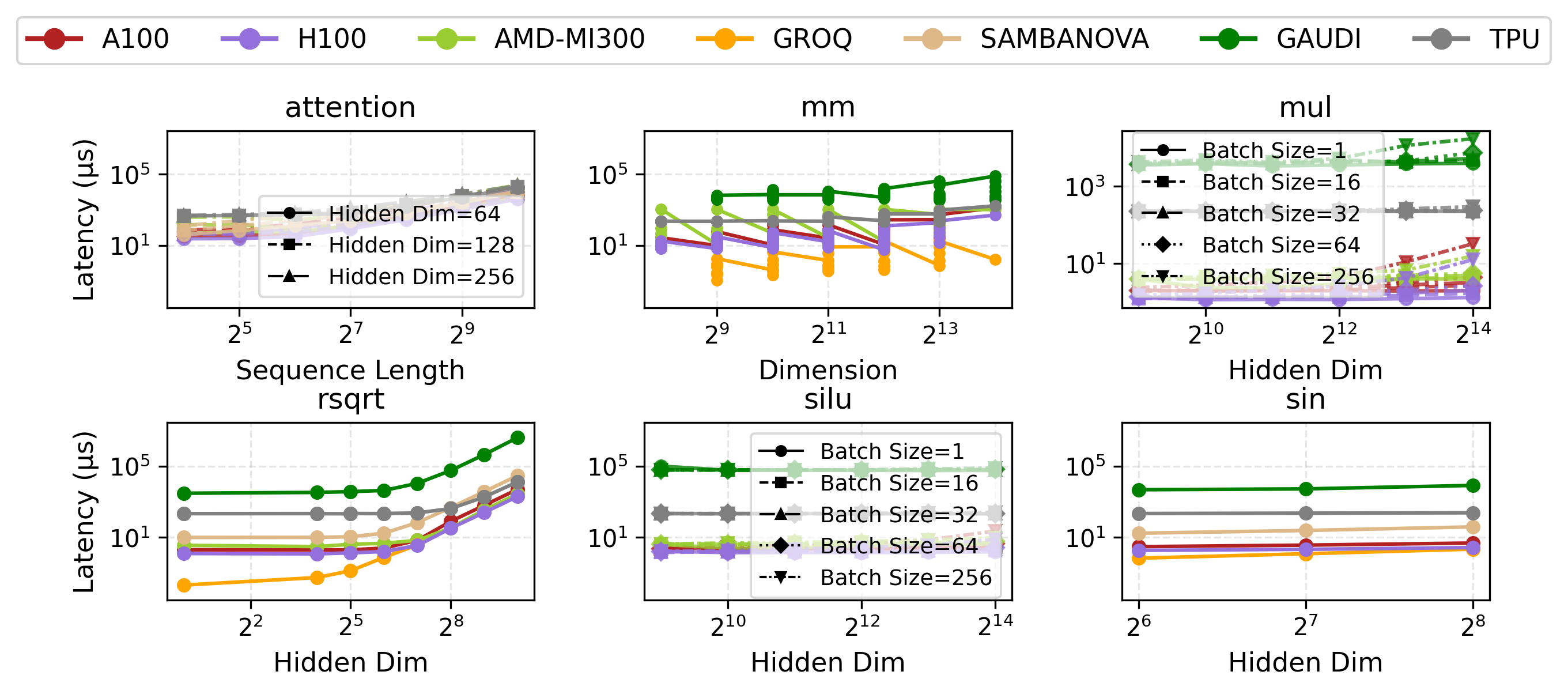}
        \caption{\textit{Latency}. Groq's advantage shines at small scale, achieves 1.64$\times$, 14.42$\times$, and 300.16$\times$ lower latency than H100 for sin, matrix multiply (mm), and rsqrt operations, respectively.}
        \label{fig:compute_benchmarks_latency_1}
    \end{subfigure}
    \hfill
    \begin{subfigure}[b]{0.49\linewidth}
        \centering
        \includegraphics[width=0.95\linewidth]{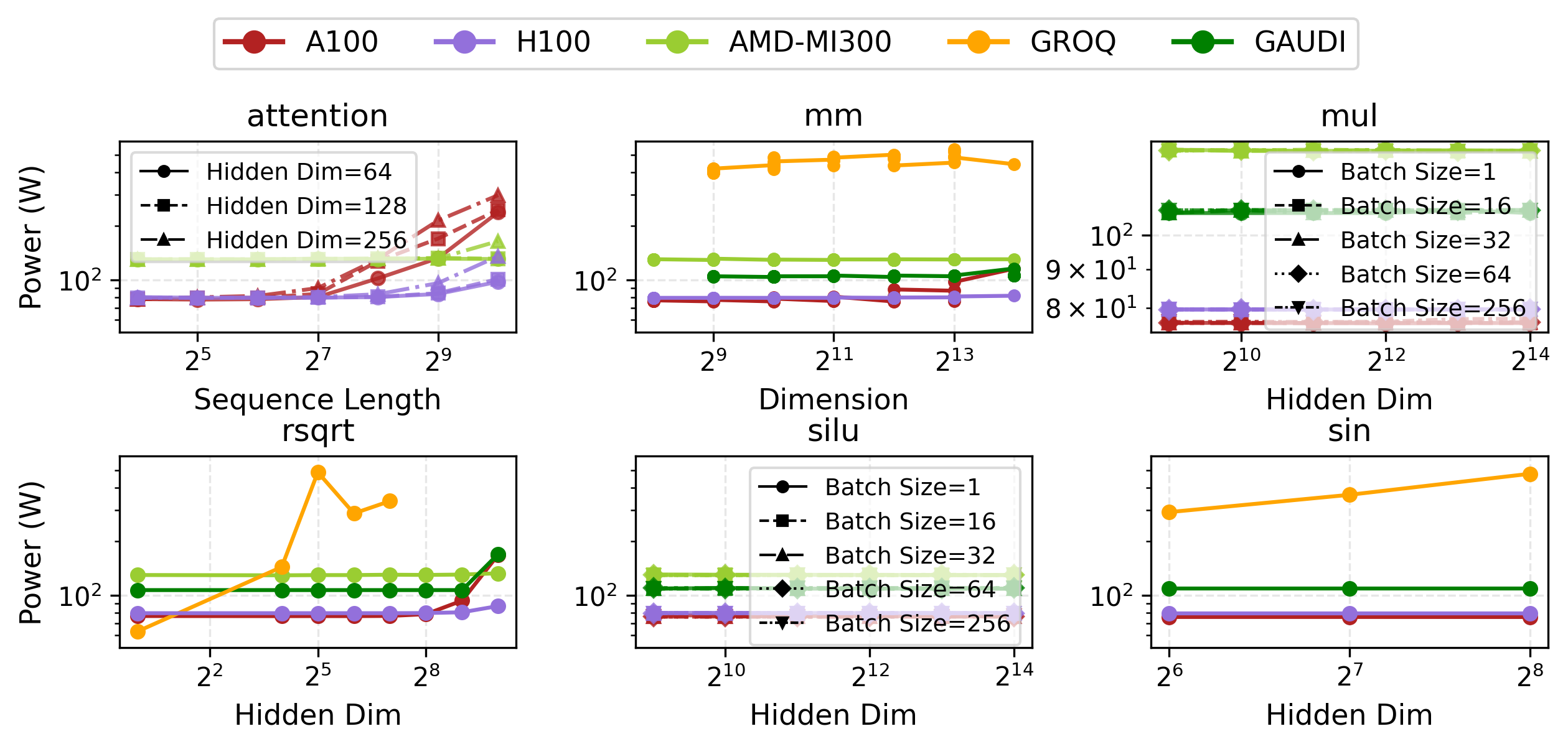}
        \caption{\textit{Power}. Even at small scales, Groq requires significant power consumption to achieve its performance, using up to 35$\times$ and 51.44$\times$ more power on rsqrt and sin operations than others, respectively.}
        \label{fig:compute_benchmarks_latency_2}
    \end{subfigure}
    \hfill
    \begin{subfigure}[b]{0.49\linewidth}
        \centering
        \includegraphics[width=\linewidth]{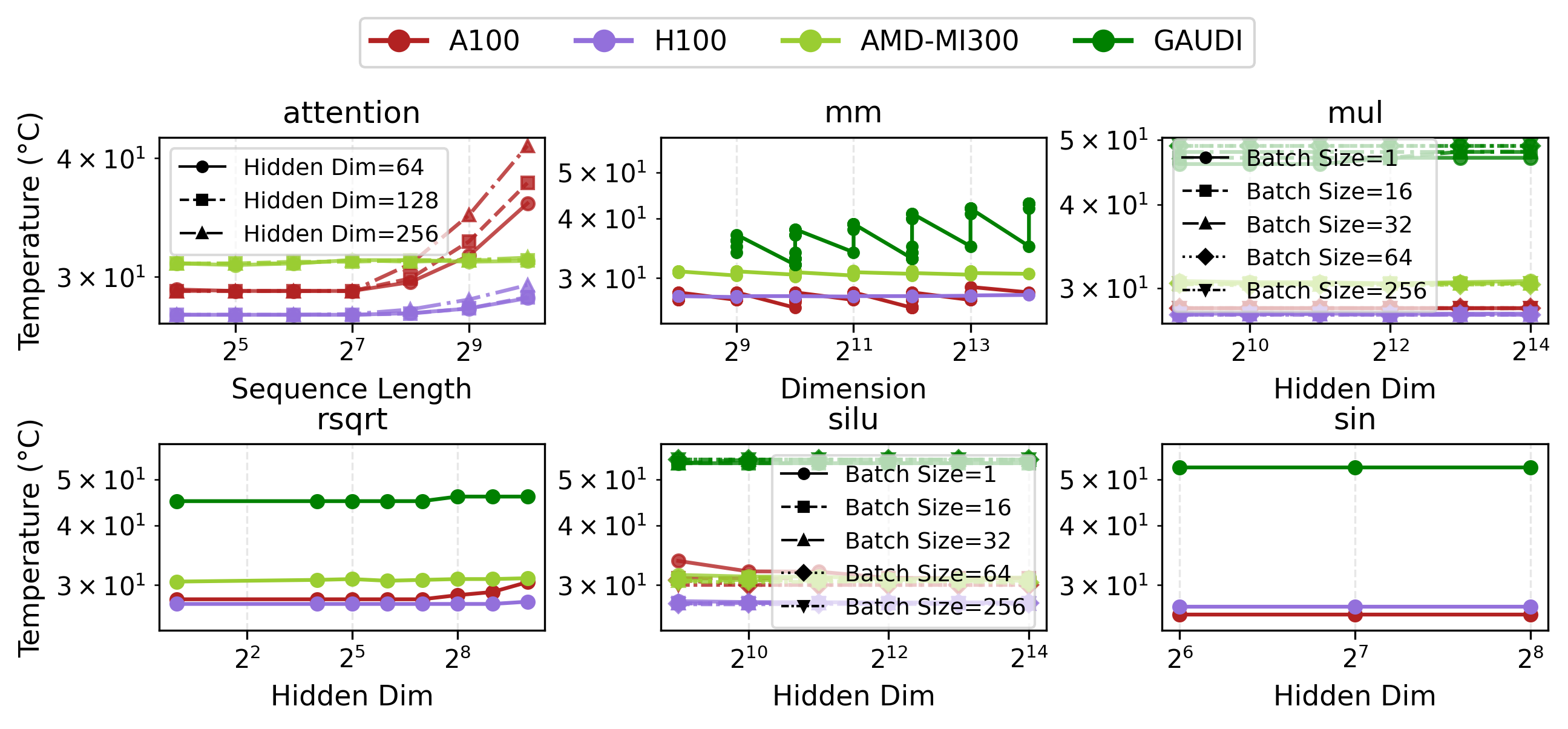}
        \caption{\textit{Temperature}. AMD-MI300 sees most consistent temperature recordings across primitives, whereas Gaudi sees the highest variation, showing different methods of temperature regulation.}
        \label{fig:compute_benchmarks_latency_3}
    \end{subfigure}
    \hfill
    \begin{subfigure}[b]{0.49\linewidth}
        \centering
        \includegraphics[width=\linewidth]{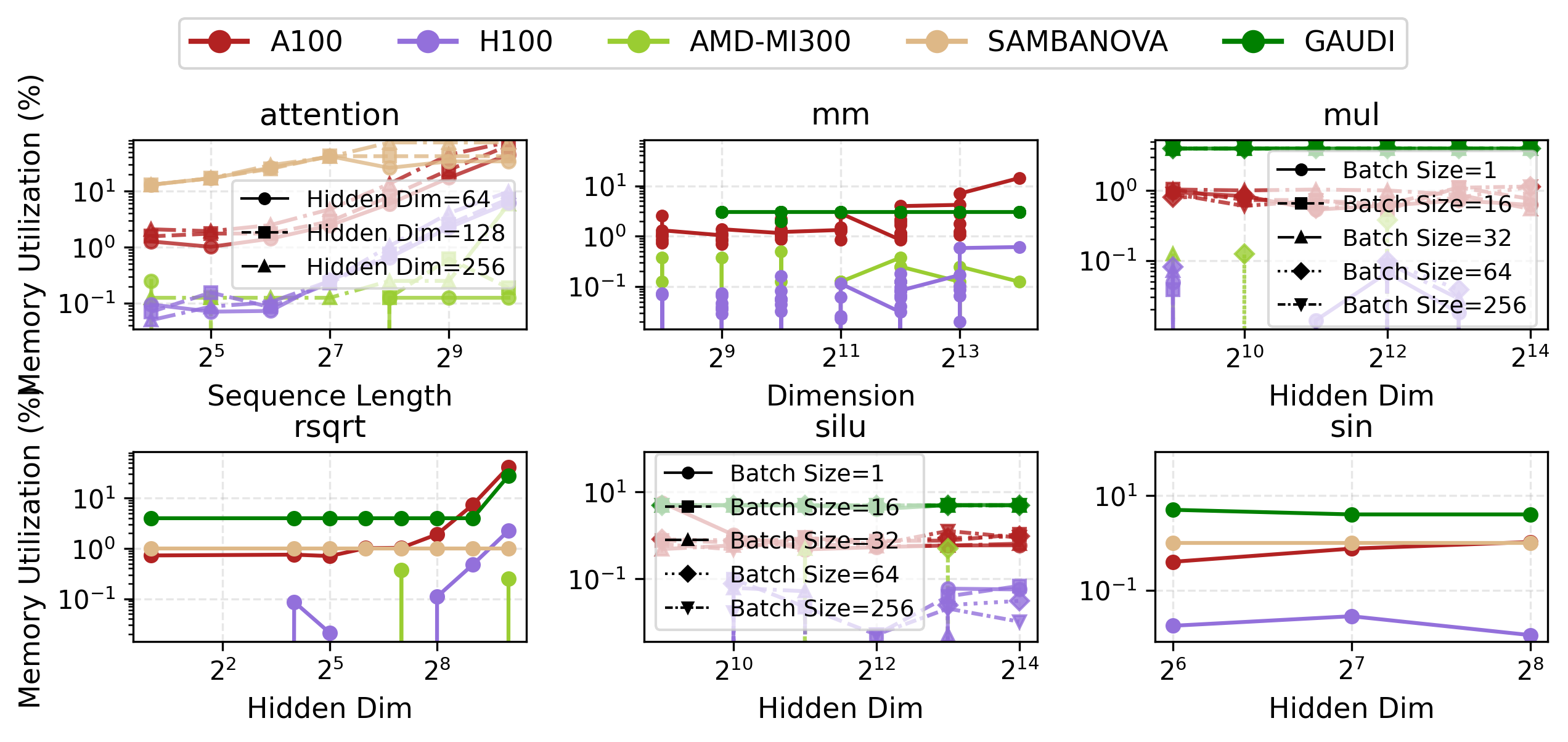}
        \caption{\textit{Memory Capacity Utilization}. Memory capacity utilization differs across architectures. The attention operation sees memory utilization range from 0.41-100\% depending on shape.}
        \label{fig:compute_benchmarks_latency_4}
    \end{subfigure}

    \caption{We highlight a subset of our computation primitive microbenchmark results here, including six operators found in LLM inference. We leverage these microbenchmarks to evaluate performance of each accelerator at small-scale.}
    \label{fig:compute_benchmarks_quad}
\end{figure*}

\section{Diving Deeper: Exploring Key Aspects of Accelerator Performance}

\subsection{Benchmarking Computational Primitives} \label{microbenchmarks}

We additionally analyze compute and communication microbenchmarks across hardware platforms in order to understand performance of computational primitives, as shown in Figure \ref{fig:compute_benchmarks_quad}. Due to space constraints, we showcase only a subset of our results here, highlighting six operators in transformer-based LLMs based on the methodology described in Section \ref{microbenchmark_methodology}. We ground our analysis on typical HuggingFace models, using the corresponding dimensions to sweep input shapes. We showcase a subset of results per operator, varying 1-2 dimensions and holding others constant.

The benchmarking results provide insights into the performance of individual operators independent of the large scale-out scenarios that must be considered for end-to-end inference. In particular, we find that on a small-scale, Groq achieves optimal performance across computational primitives, and is 1.64x faster than H100 for sin operations, 14.42$\times$ faster for matrix multiply (mm), and 300.16$\times$ faster for rsqrt operation. We note, however, that Groq's speed advantage is limited to operators that it supports, and it notably lacks support for pytorch scaled dot product attention (SDPA) kernel. 

\vspace{0.5em}
\noindent
\fcolorbox{black!100}{gray!10}{
  \parbox{0.95\linewidth}{
    \vspace{0.2em}
    \underline{Key Takeaway 5:} Groq’s latency advantage emerges primarily at small scales, as its strength comes from extremely fast computational primitives rather than from large scale-out deployments. 
    \vspace{0.2em}
  }
}
\vspace{0.5em}

We additionally use these benchmarks to highlight SambaNova’s strengths in accelerating specific kernels. Unlike Groq—which lacks SDPA support and must rely on non-fused attention—SambaNova performs well with fused kernels. In fact, SambaNova outperforms A100 on large fused ops (e.g., SDPA, where it is 1.12× faster) but lags on general-purpose ops like rsqrt and sin, where A100 can be up to 7× faster.

We record power, temperature, memory usage, clock frequency, and memory utilization across benchmarks for all supported platforms. Groq is the most power-hungry platform, drawing up to 35× more power than others on rqsrt, 51.44× more on sin, and 18.61× more on matrix multiply. Temperature regulation also varies: AMD MI300 maintains nearly constant temperatures, whereas Gaudi shows the most dynamic thermal behavior. Memory capacity utilization similarly differs by platform, with attention workloads varying 0.41--100\%.

\subsection{Energy Cost of Communication.}

To better understand communication costs of AI accelerators in distributed inference settings, we quantify the communication energy required to send data a given distance. To do so, we design microbenchmarks that capture the transfer of $B$ bytes of data between silicon distances. We profile power and execution time, and calculate energy as shown:
\[energy=
\frac{\,P_{\text{benchmark}} - P_{\text{idle}}\,}{time}
\]
where $P_{benchmark}$ and $P_{idle}$ represent the combined power consumption of the platforms involved in the system, i.e., both chips if transferring across two accelerators.

\begin{figure}[t]
    \centering

    \begin{subfigure}[t]{0.38\linewidth}
        \centering
        \includegraphics[width=\linewidth]{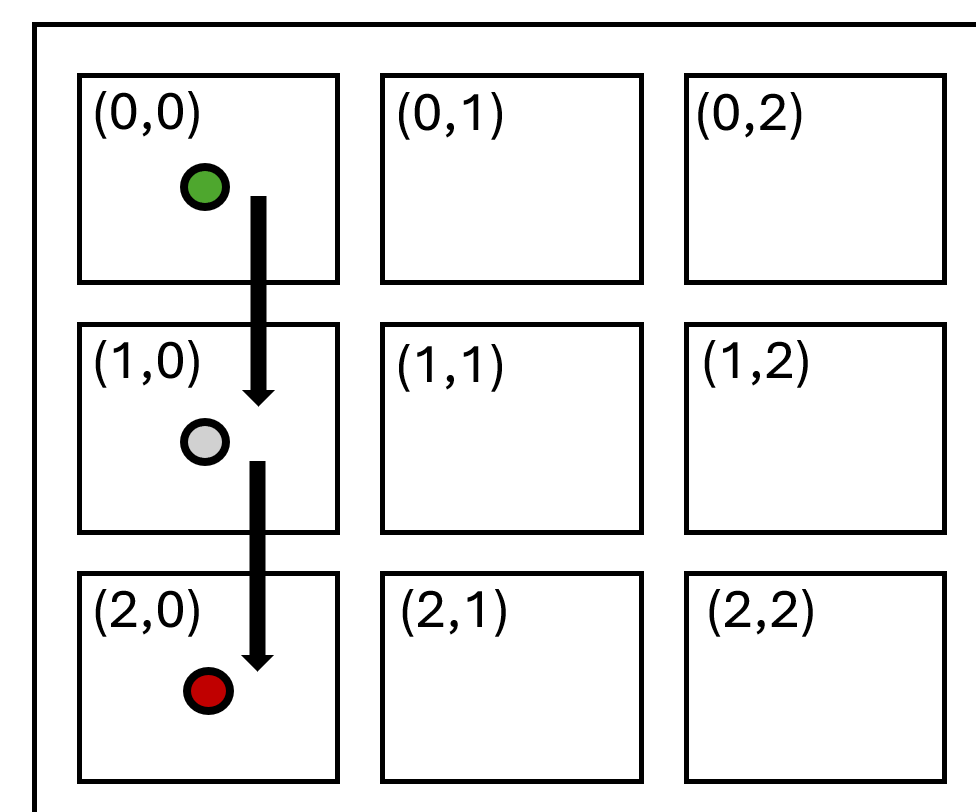}
        \caption{CS-3 benchmark leverages SDK to send data from sender PE to receiver PE at varying distances.}
        \label{fig:comm_1}
    \end{subfigure}
    \hfill
    \begin{subfigure}[t]{0.58\linewidth}
        \centering
        \includegraphics[width=\linewidth]{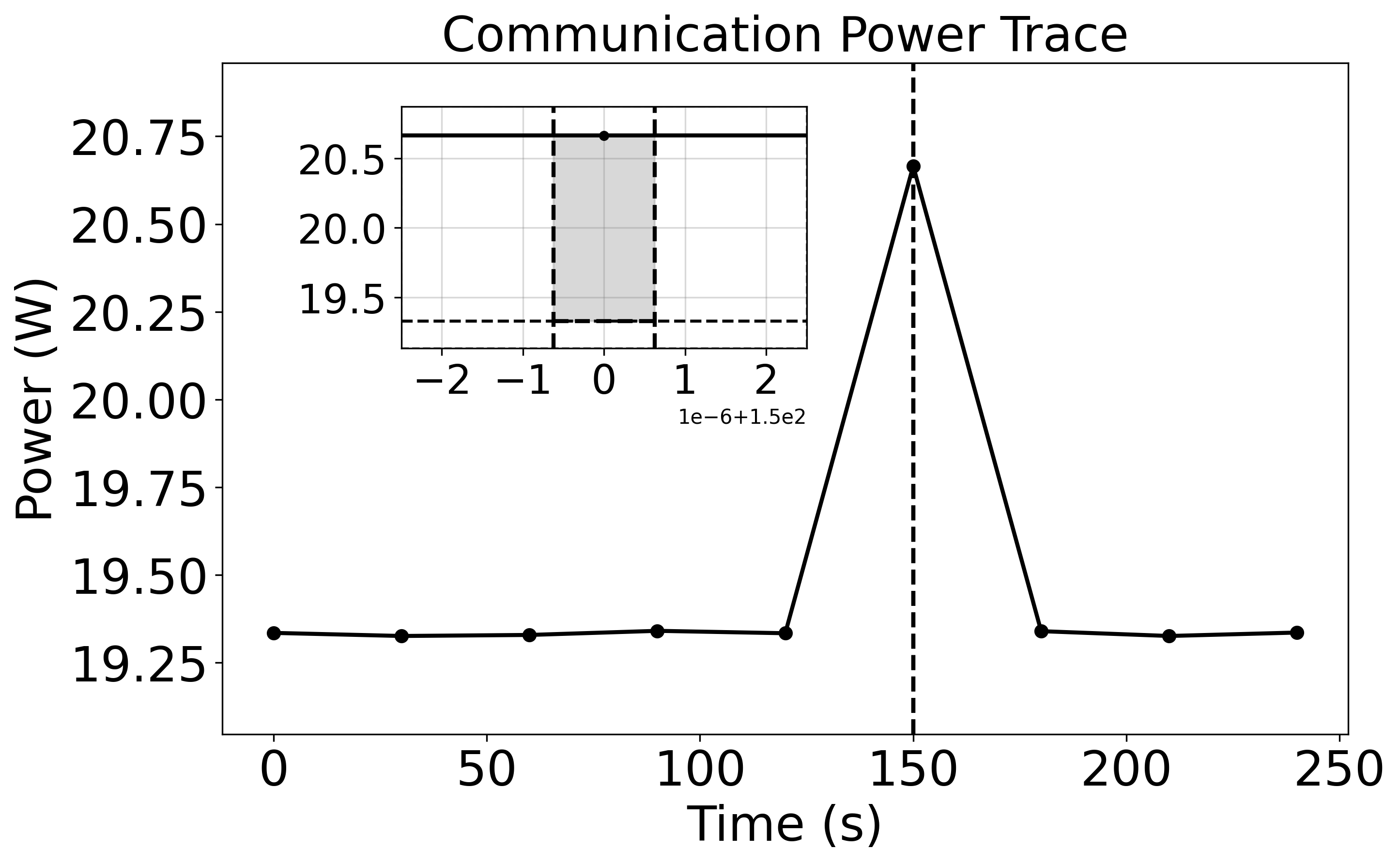}
        \caption{Communication energy is quantified by capturing power trace during data transfer, subtracting out idle power, and multiplying by execution time.}
        \label{fig:comm_2}
    \end{subfigure}


    \begin{subfigure}[t]{0.37\linewidth}
        \centering
        \includegraphics[width=\linewidth]{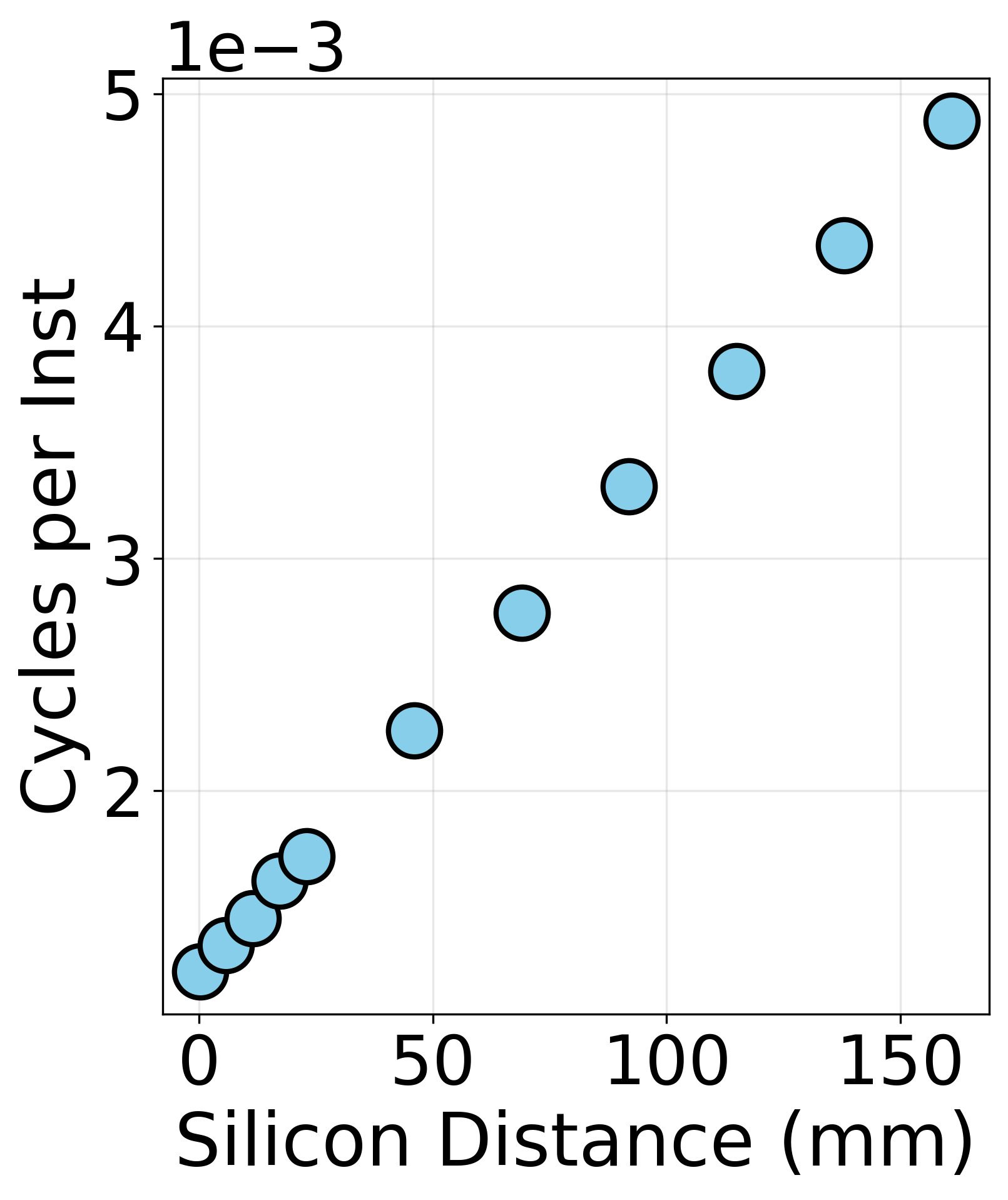}
        \caption{Cycles per data transfer instruction, as captured on CS-3 wafer.}
        \label{fig:comm_3}
    \end{subfigure}
    \hfill
    \begin{subfigure}[t]{0.58\linewidth}
        \centering
        \includegraphics[width=\linewidth]{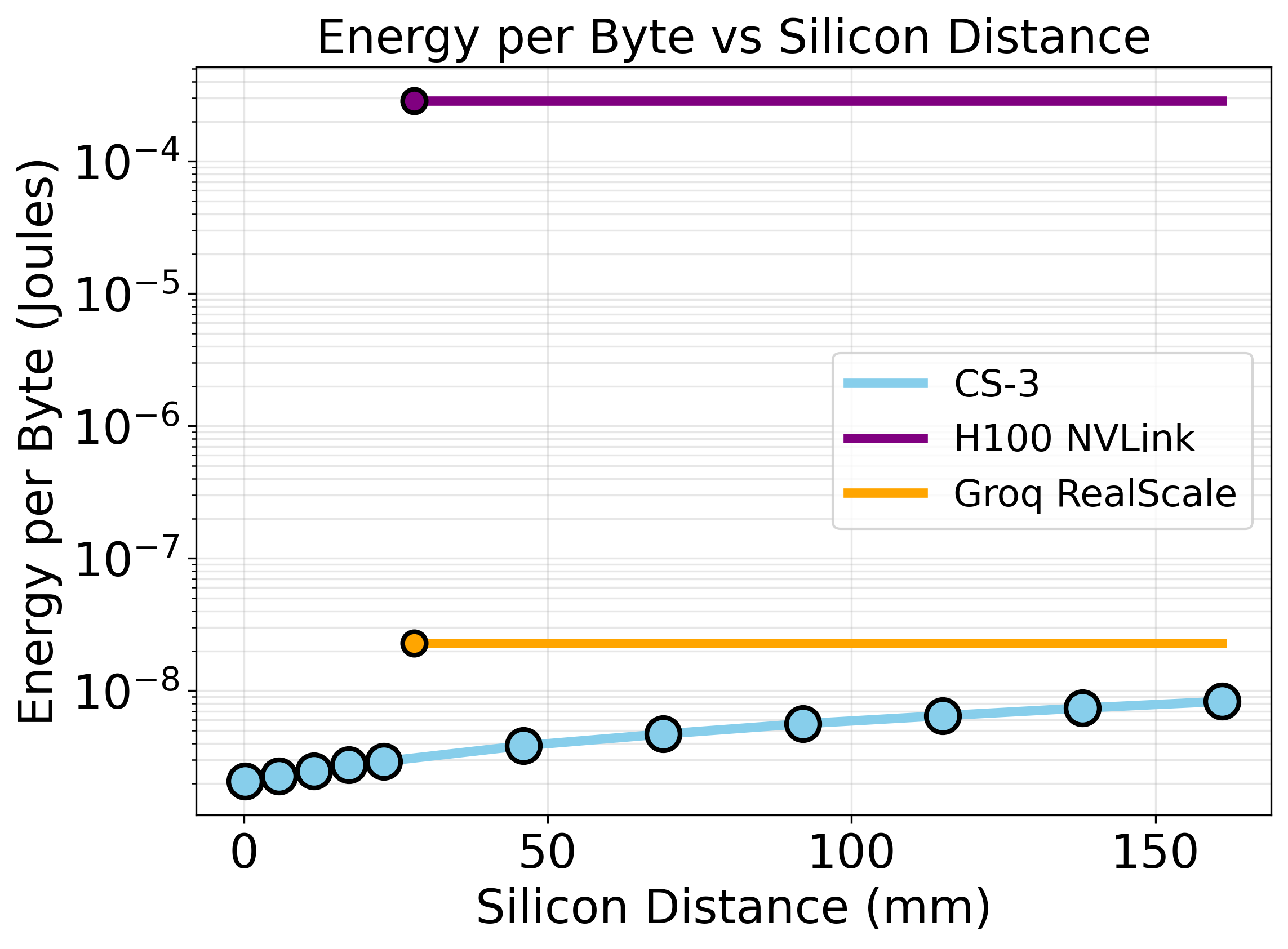}
        \caption{Communication energy comparison. We find CS-3 uses up to 74,433$\times$ less energy per byte than H100.}
        \label{fig:comm_4}
    \end{subfigure}

    \caption{Communication energy comparison across Cerebras, Groq, and H100. We find Cerebras CS-3 is significantly more efficient at transferring data than traditional GPUs.}
    \label{fig:comm}
\end{figure}

We first measure the energy cost of communicating for CS-3. We design a microbenchmark to send data across PEs within a wafer. To do so, we utilize the Cerebras SDK and low-level CSL programming language to specify a sender PE and receiver PE and manually orchestrate the data transfer between them, as shown in Figure \ref{fig:comm_1}. We then leverage the SDK simulation tool to verify and trace the expected data transfers (i.e., wavelets). Finally, we run our benchmark on the physical wafer, and analyze PDU output to quantify power consumption. Figure \ref{fig:comm_2} shows an example power trace from our CS-3 benchmark, where we quantify energy based on formula (1) above. We sweep sender and receiver PEs to quantify the energy cost at different distances on wafer. Figure \ref{fig:comm_3} shows how cycles scale with silicon distance.

We then compare this to the energy needed to communicate across Groq chips. To quantify this, we design a microbenchmark to send data across chips, and verify the transfer through power utilization of the Chip-to-Chip (C2C) unit. We follow the methodology above to calculate energy, quantifying the associated cycles and power when the C2C unit is activated. We ground our comparisons in an H100 baseline, following a similar methodology to transfer data via NVLink. As shown in Figure \ref{fig:comm_4}, we find that when communicating over distances of 161mm, CS-3 uses 34,454$\times$ less Joules/byte than an H100 system, and 2.74$\times$ less energy per byte than Groq.

\vspace{0.5em}
\noindent
\fcolorbox{black!100}{gray!10}{
  \parbox{0.95\linewidth}{
    \vspace{0.2em}
    \underline{Key Takeaway 6:} Cerebras significantly reduces the energy cost of communication as compared to H100 and Groq when models fit inside a single wafer, since the large silicon area allows for data movement to stay on-chip. 
  }
}
\vspace{0.5em}

\section{Discussion}

\subsection{Challenges of Accelerator Profiling}

\begin{figure}[b]
    \centering
    \begin{subfigure}[b]{0.45\linewidth}
        \centering
        \includegraphics[width=\linewidth]{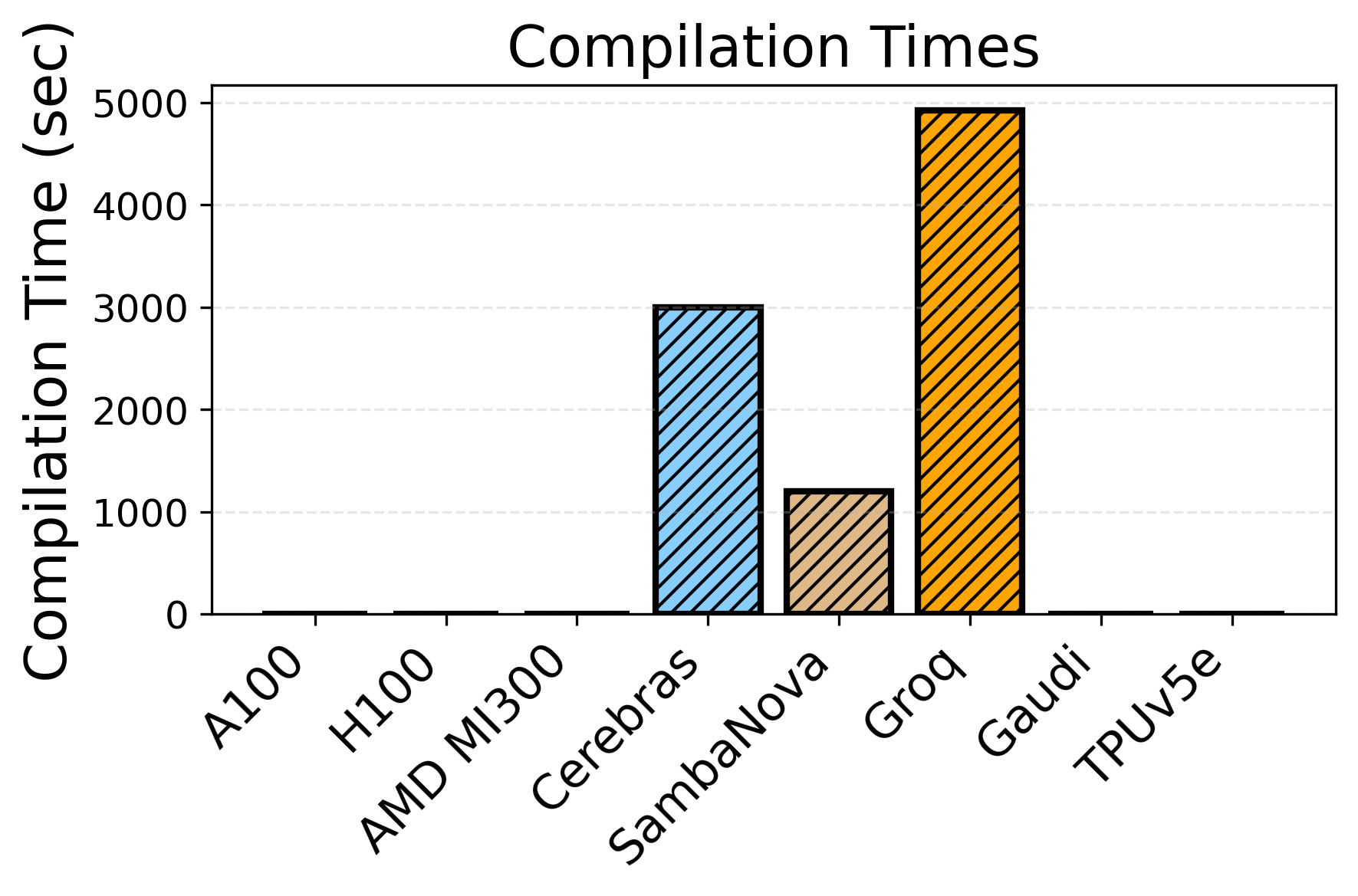}
        \caption{Build time for typical LLM inference. Groq has longest compilation times, followed by Cerebras and SambaNova.}
        \label{fig:compute_benchmarks_latency_1}
    \end{subfigure}
    \hfill
    \begin{subfigure}[b]{0.53\linewidth}
        \centering
        \includegraphics[width=\linewidth]{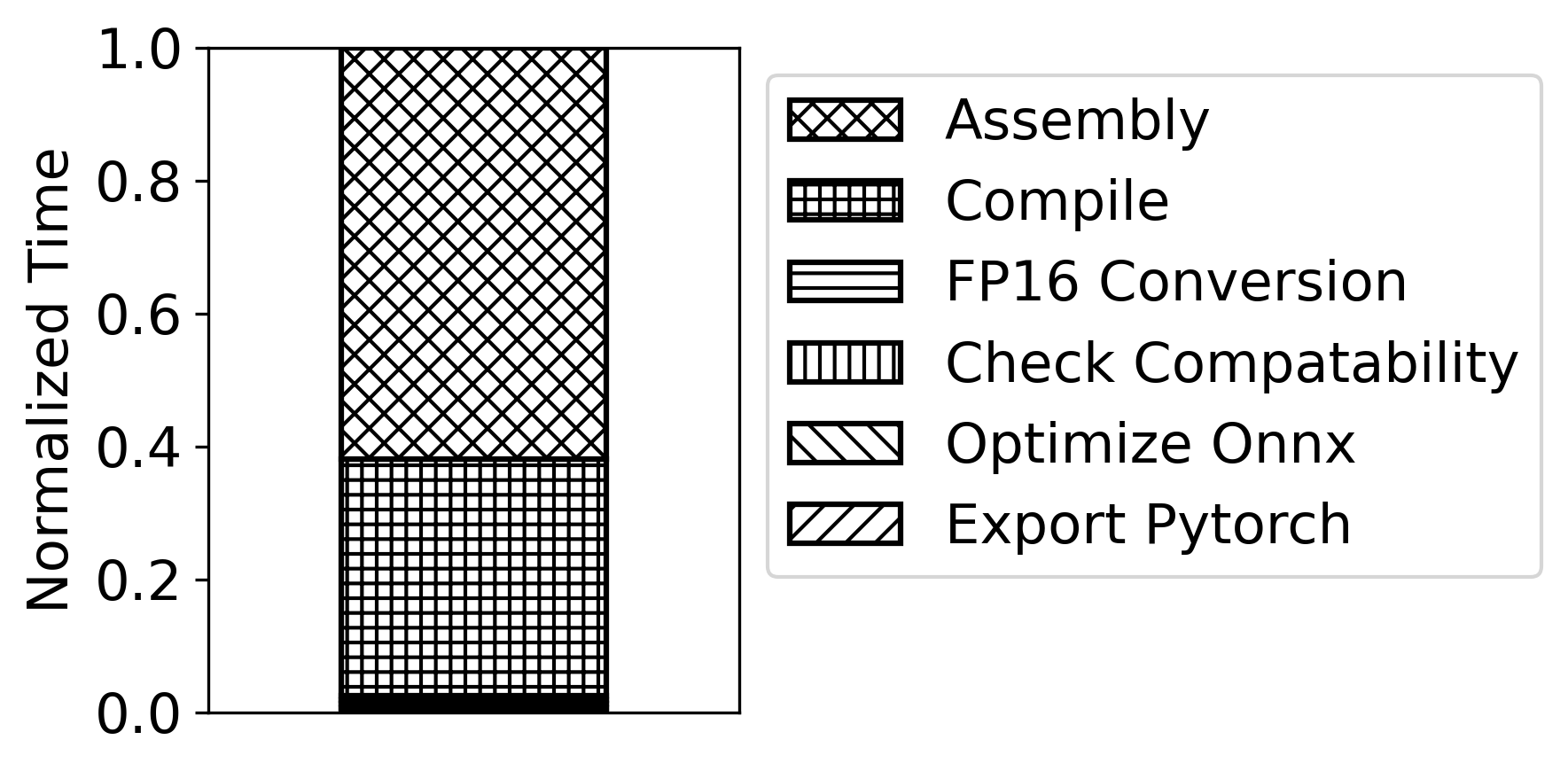}
        \caption{Detailed view of Groq build time. We find compilation and assembly stages dominate, consuming 40\% and 59\% of total build time. }
        \label{fig:compute_benchmarks_latency_3}
    \end{subfigure}
    \\
    
    \caption{Groq sees largest build times, followed by Cerebras and SambaNova. Groq also requires highest build frequency.}
    \label{fig:compilation}
\end{figure}

Throughout our evaluation of each accelerator platform, we found the most significant challenge resulted from getting models to run effectively. In this section, we share some lessons learned with the community.

\begin{itemize}

    \item \textbf{Programmability.} Each accelerator requires their own set of toolchains, compiler configurations, and profiling workflows. While all platforms boast easy-to-use model zoos, repos are unfortunately often incomplete and caveats abound on running the advertised models. For example, even simple batch size scaling is not supported with our physical CS-3 and Groq setup. Inconsistent software, undocumented limitations, and incomplete kernel coverage often stalled progress or forced workarounds. 

    \item \textbf{Compilation and Build Times.} Specialized accelerators often have large build times, which can hinder the iterative research process. Figure \ref{fig:compilation} highlights the respective build times across accelerators. We observe build times for Groq can be upwards of 4500 seconds, and scale linearly with the number of chips. Note the significance of build time differs depending on use-case and how frequently parameters change.
    
    \item \textbf{Not All Metrics Are Created Equal.} Even the same metrics can have different definitions across hardware platforms. For example, NVIDIA utilization refers to SM utilization, while Cerebras refers to compute flops. 

    \item \textbf{Sudo Access / Admin Privileges.} Limited administrative privileges can constrain profiling. In order to collect power on Cerebras and SambaNova, for example, we had to partner with cluster administrators who had sudo privileges to gather PDU data.
    
\end{itemize}

\subsection{Comparing Programmability}
Here we examine programmability in greater depth, providing a detailed analysis of each accelerator’s software stack. We compare each platform along key dimensions that capture software programmability including tooling maturity and kernel support, in addition to the compilation analysis shown.

PyTorch support varies significantly across accelerator platforms, and is often less straightforward outside mainstream ecosystems. For example, Cerebras supports CSTorch, which is a specialized PyTorch API designed to directly run on the Cerebras wafer. However, there are limitations, including tensors that must be initialized ahead of time, and a special step closure is needed to prevent retrieving tensor values prematurely. Additionally, custom template code is required to run individual operations on the Cerebras system, and it is not currently possible to run individual operator benchmarks.

In a similar manner, SambaNova, like Cerebras, requires a specialized PyTorch language designed for SambaNova to run ordinary PyTorch operations. SambaFlow is a specialized Python SDK used for programming on SambaNova. As compared to the 3000+ PyTorch operators, SambaFlow supports an operator count in the low hundreds. Tensors must be defined with the samba package and common distributed communication primitives, are provided through SambaFlow-specific implementation.

In contrast, while Groq's support of the PyTorch language is more direct, the set of PyTorch operators supported is quite limited, supporting only a few hundred. Groq notably lacks support for several key operators when running on their physical hardware platform, including torch.mm and the standard PyTorch attention operation (i.e., Scaled Dot Product Attention). Furthermore, code wrappers are needed to export code and convert to onnx before it can be run on the hardware. Functions must follow a uniform in-house template, and while running outside this template may be theoretically possible, there is no public documentation or support.

Additionally, although model zoos advertise broad support, running these models in practice can be surprisingly difficult. Note that we evaluate this in the context of running the models on physical hardware, not leveraging the online API. For example, despite advertised LLaMA support, deploying a LLaMA model on nine Groq racks required approximately 1.5 years of coordination and repeated compiler issues, and ultimately only hand-compiled models from Groq engineers were executable; variants could not be run. In contrast, Cerebras model zoo supports over 30 workloads, though its open-source tooling primarily targets training. SambaNova also provides state-of-the-art models, albeit with configuration limitations.

\subsection{Call for Software Stack Enablement.}

Poor software enablement severely limits practical usability. Real performance depends on how effectively compilers, runtimes, and kernel libraries target the hardware; missing ops and immature tooling can hide true capabilities. To realize the full potential of AI accelerators, the community must focus efforts on developing the software stack --- in particular, advanced compilation flows and optimized kernel development. We note LLMs introduce a new opportunity to minimize this software overhead, generating optimized, customized kernels in a fraction of the time of human engineers. We believe this is an important area of work for the community to invest, such that the full potential of these accelerators can be realized.

\section{Related Work}

Many existing hardware characterization frameworks, such as MLPerf and MLPerf Power~\cite{mlperfInference} focus on conventional CPU and GPU architectures and offer only limited support for specialized AI accelerators. MLPerf standardizes benchmarking for CPUs, GPUs, and recently some TPUs. Recent work characterizing more diverse AI accelerators has tended to emphasize high-level ML metrics over architectural analysis. Platforms like ArtificialAnalysis~\cite{artificialanalysis2025hardware} benchmark accelerators primarily through aggregate throughput, offering useful comparisons but limited insight into underlying system properties. Other studies, such as those examining edge AI accelerators~\cite{intelligencePerWatt}, report performance-per-watt using coarse estimates like TDP and measure performance via high-level APIs, capturing overall efficiency but not operator-level behavior or real power draw under dynamic workloads. Other studies such as~\cite{argonnePaper} evaluate novel accelerators, but only quantify throughput and not other system properties. \cite{argonnePaper} also focuses on older transformer models and is geared towards training, not LLM inference.

\section{Conclusion}

Through our quantitative evaluation across eight AI accelerators, we showcase how the architectural properties translate into real performance and efficiency trade-offs for end-to-end LLM inference workloads. These findings underscore the need for hardware-algorithm co-design and robust software stacks in order to enable next-generation heterogeneous systems.

\section{Acknowledgments}

The authors at Harvard University are supported in part by the National Artificial Intelligence Research Resource (NAIRR) Pilot and the Argonne National Lab AI Testbed at the Argonne Leadership Computing Facility, a U.S. Department of Energy (DOE) Office of Science user facility at Argonne National Laboratory based on research supported by the U.S. DOE Office of Science–Advanced Scientific Computing Research Program under Contract No. DE-AC02-06CH11357; the Neocortex system supported by the National Science Foundation (award NSF-OAC 2005597) at the Pittsburgh Supercomputing Center, a joint computational research center with Carnegie Mellon University and the University of Pittsburgh; and the National Science Foundation Expedition in Computing (CCF-2326605, 2326606, 2326607, 2326608, 2326609, 2326610, and 2326611). We note that Alicia led the experimental work for this paper, and Carole-Jean Wu, Gu-Yeon Wei, and David Brooks were advisors. Any opinions, findings, conclusions, or recommendations expressed in this material are those of the authors and do not necessarily reflect the views of this sponsor.

\newpage

\bibliographystyle{IEEEtranS}
\bibliography{references}
\end{document}